\documentclass[traditabstract]{aa} 
\usepackage{txfonts}
\usepackage{longtable}

\usepackage[dvips]{graphicx}
\usepackage{natbib}
\usepackage{tabls}
\usepackage[format=hang,font=small,labelfont=bf]{caption}
\bibpunct{(}{)}{;}{a}{}{,} 
\usepackage{subfig}
\usepackage{ctable}
\newcommand*{\ms}{\ensuremath{\mathrm{M^{*}}}}
\newcommand*{\Msun}{\ensuremath{\mathrm{M_\odot}\,\,}}%
\begin{document}
   \title{The Galaxy Stellar Mass Function of X--ray detected groups}

   \subtitle{Environmental dependence of galaxy evolution in the COSMOS survey}

   \author{S. Giodini \inst{1}	 
    \and A. Finoguenov \inst{2}
	\and D.Pierini \thanks{Visiting astronomer at Max-Planck-Institut fuer extraterrestrische Physik, Giessenbachstrasse 1, D-85748 Garching, Germany}
	\and G. Zamorani \inst{3}
	\and O. Ilbert \inst{4}		
	\and S. Lilly \inst{5}
	\and Y. Peng \inst{5}
	\and N. Scoville  \inst{6}	
	\and M. Tanaka \inst{7}		
	}         

   \institute{Leiden Observatory, Leiden University, PO Box 9513, 2300 RA Leiden, the Netherlands \\
              \email{giodini@strw.leidenuniv.nl}
         \and
              Max-Planck-Institut fuer extraterrestrische Physik, Giessenbachstrasse 1, D-85748 Garching, Germany
		\and				 
             	INAF-Osservatorio Astronomico di Bologna, Via Ranzani 1, I-40127 Bologna, Italy	              		
        \and
        		 Laboratoire d’Astrophysique de Marseille, BP 8, Traverse du Siphon, 13376 Marseille Cedex 12, France   		
        	\and
        		Department of Physics, ETH Z{\"u} rich, CH-8093 Z{\"u}rich, Switzerland
       \and				
        	   California Institute of Technology, MS 249-17 Pasadena, CA 91125, USA
		\and        		
        		Institute for the Physics and Mathematics of the Universe, University of Tokyo, Kashiwa 2778582, Japan       				
            }

   \date{Received --; accepted --}

 
  \abstract
{ We study the stellar mass distribution for galaxies in 160 X-ray detected groups of  10$^{13}<$Log(M$_\mathrm{200}$/M$_{\odot}$)$<$2$\times$10$^{14}$ and compare it with that of galaxies in the field, to investigate the action of environment on the build up of the stellar mass. We highlight differences in the build up of the passive population in the field, which imprint features in the distribution of stellar mass of passive galaxies at Log(M/M$_{\odot}$)$<$ 10.5. The gradual diminishing of the effect when moving to groups of increasing total masses indicates that the growing influence of the environment in bound structures is responsible for the build up of a quenched component at Log(M/M$_{\odot}$)$<$ 10.5. Differently, the stellar mass distribution of star forming galaxies is similar in shape in all the environments, and can be described by a single Schechter function both in groups and in the field. Little evolution is seen up to redshift 1. Nevertheless at z=0.2--0.4 groups with M$_{200}<$6$\times$10$^{13}$\Msun (low mass groups) tend to have a characteristic mass for star forming galaxies which is 50$\%$ higher than in higher mass groups; we interpret it as a reduced action of environmental processes in such systems. \\
Furthermore we analyse the distribution of sSFR--Log(M) in groups and in the field, and find that groups show on average a lower sSFR (by $\sim$0.2 dex) at z$<$0.8. Accordingly, we find that the fraction of star forming galaxies is increasing with redshift in all environments, but at a faster pace in the denser ones.  \\
Finally our analysis highlights that low mass groups have a higher fraction (by 50$\%$) of the stellar mass locked in star forming galaxies than higher mass systems (i.e. 2/3 of their stellar mass).}

   \maketitle
%

\section{Introduction}

   The galaxy stellar mass function (GSMF) is a very important diagnostic to perform a census of galaxy properties, and provides powerful means of comparison between the populations of galaxies in different environments.\\
Historically, the luminosity function has been the first diagnostic used to study the distribution of galaxy properties, since a magnitude is a more direct observable than mass (Schechter 1976; Binggeli et al. 1988). However, the development of stellar population synthesis models and deep multi--wavelength surveys have greatly improved our ability to estimate the stellar mass content in galaxies. We can now study the distribution in stellar mass, a parameter which is more directly linked to the total mass of a galaxy. \\
The galaxy stellar mass function is important for both cosmology and galaxy evolution to better understand the connection between galaxy and dark matter distributions, and their link to the environment. In particular, the shape of the GSMF and its evolution give very important insights into the processes that contribute to the growth in stellar mass of galaxies with time and that drive the formation and evolution of galaxies in different environments. \\
The GSMF has been extensively studied  in deep fields, for galaxies of different colors and morphological types \citep{2006ApJ...651..120B, Baldry:2008p58, 2010A&A...523A..13P} and in different environments \citep{Balogh:2001p86, Yang:2009p343, Vulcani:2010p919}. Its shape has been described by  a Schechter function \citep{Schechter:1976p1608}, that is an empirical model also used to describe the luminosity function.  When fitted to the data, the shape of this function changes both as a function of the galaxy type (star--forming/passive, or morphological type) and of the environment \citep{Balogh:2001p86, Bolzonella:2009p425, Yang:2009p343}.\\
The low-mass end of the galaxy stellar mass function is an important constrain for galaxy formation models, which generally overpredict the observed number of dwarf galaxies \citep{2011arXiv1105.0674W}.
The availability of deeper optical and infrared data enabled a study of the low mass end  of the GSMF, showing a more complicated behaviour than a single Schechter function. This result was already suggested by luminosity function studies,  where an excess of faint galaxies has been revealed both in  deep fields \citep{Baldry:2004p4667, Trentham:2002p4697, Blanton:2005p4751} and in studies focused on galaxy clusters and groups \citep{Wilson:1997p4567, Hilton:2005p4608, Popesso:2005p1365, Gonzalez:2006p4487}. 
In the light of these results, different authors suggested that the GSMF may be better described by a multicomponent model, such as a double power law \citep{Yang:2009p343} or a double Schechter function obtained by adding a second Schechter function, with a steep negative slope and a lower characteristic mass \citep{Driver:1994p1615, Baldry:2008p58, Drory:2009p22}.Thereby the first term, $\phi_{1}$(M), is identified with a population of massive galaxies and the second term, $\phi_{2}$(M), with a population of low mass galaxies.\\
From a theoretical point of view, the shape of the Schechter function (characterized by a slope and a characteristic mass) calls for a physical interpretation.
On one hand, a mass function with a steep rising slope at low stellar masses is a generic prediction of CDM models \citep{White:1978p1621}, if galaxies follow the underlying halos and sub-halos mass distribution. Differences between the shape of the galaxies and dark matter mass distribution are likely driven by non-gravitational processes connected with star--formation and feedback in galaxies; therefore the slope of the GSMF is an important constraint for the modelling of non gravitational processes. 
On the other hand, the characteristic mass which defines the knee of the GSMF (\ms) is interpreted as a threshold where galaxy growth by star formation is not an efficient process, and is overruled by growth through merging processes \citep{2009MNRAS.397..506K}.  Moreover, simulations have shown that the steep cut-off at high stellar masses can be reproduced by taking into account feedback from supermassive black holes as a main ingredients in galaxy evolution\citep{2006MNRAS.365...11C}.\\
With the advent of large multi-wavelength surveys, data achieved a sufficient enough accuracy to provide a guideline for galaxy evolution models. 
\citet{Peng:2010p568} demonstrated the fruitfulness of a data--based approach: starting from observed properties of the galaxy distribution in SDSS DR7 and zCOSMOS, these authors devise a simple description of how star formation is quenched in the global galaxy population.  When this model is applied to a simulated sample of galaxies, it correctly reproduces the observed GSMF in the global field.\\
Since a large fraction of the universal stellar mass is formed in galaxy groups (M$_\mathrm{tot}<$10$^{14}$ \Msun ; \citealt{Crain:2009p1881}), it is crucial to study the GSMF in these environments to have a complete understanding of the mass assembly. 
Another intriguing aspect of studying galaxy groups is the compelling evidence that  most of the pre-processing of galaxies occurs in groups'-sized halos \citep{vandenBosch:2008p3931, Wetzel:2011p3934} before falling in more massive structures.
Furthermore, groups of galaxies exhibit a correlation between the baryon fraction locked in stars and the group mass \citep{Giodini:2009p923, Mcgaugh:1p2899}, suggesting that low mass systems are the most effective environment for the conversion of baryons into stars. Our aim is to use the GSMF as a tool to shed light  on this phenomenon and constrain the stellar mass content of these systems.\\
Deep X--ray surveys, as the one performed on COSMOS \citep{Scoville:2007p1888, Hasinger:2007p1957}, provide for the first time enough information to perform a statistical study on a large  sample of X--ray selected galaxy groups.\\
We investigate the GSMF of the X--ray selected groups in the COSMOS 2 degs$^2$  field and compare it to that of clusters and the field. The COSMOS survey provides a unique database of photometric and spectroscopic data, together with deep X--ray data from XMM and Chandra, and the largest catalog of X--ray detected groups up to now. 
We take advantage of the X--ray data to provide a definition of environments based on the depth of the dark matter potential well, dividing between low mass and high mass groups. Furthermore, X-ray information provides evidence for a  gravitationally bound nature of the identified groups and a better total mass proxy, giving a more solid basis for subsequent conclusions. \\
 This paper is structured as follows: in section \ref{samples}  we describe the sample of X--ray detected groups (\ref{groups}) and the sample of group member galaxies (in \ref{galaxies} ). In section \ref{results} we present and analyze the GSMF for the COSMOS X--ray selected galaxy groups, comparing it with that of the field; in section \ref{baryon_fraction} we study the fraction of baryons in galaxies in high and low mass groups and in section \ref{check_sf} we compare the stellar mass fractions obtained in this work with other values in the literature. Finally in section \ref{ssfr} we study the distribution of specific star formation rate in the different environments. Results are discussed in section \ref{discussion}.\\
 We adopt a $\Lambda$CDM cosmological model ($\Omega_{m}$ = 0.27, $\Omega_{\Lambda}$ = 0.73) with H$_{0}$ = 71 km s$^{-1}$ Mpc$^{-1}$.

\section{The sample}
\label{samples}
\subsection{Galaxy groups in the COSMOS survey}
\label{groups}
\begin{figure}
\begin{center}
\includegraphics[width=\columnwidth]{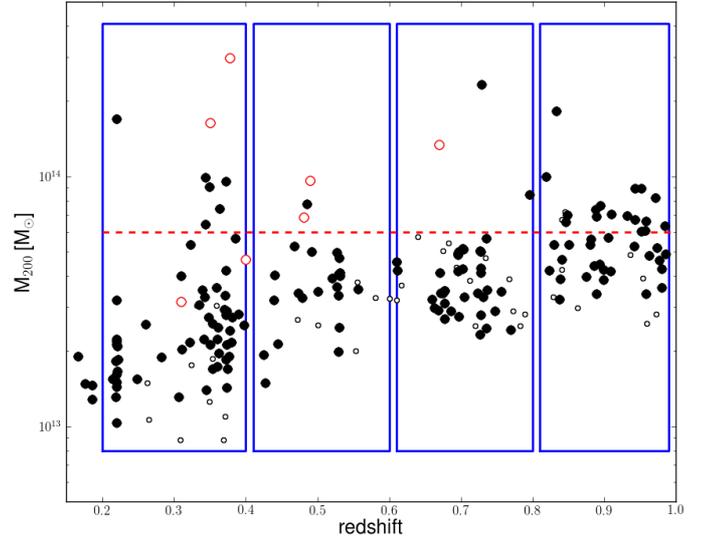}
\caption{Distribution of M$_{200}$ as a function of redshift for the X--ray detected groups in the COSMOS survey up to z=1.0. The black filled circles are the COSMOS groups used in this work. Red open circles mark the systems excluded from the analysis because out of the SUBARU area. Empty circles mark low significance and low quality groups excluded from the analysis. Rectangles show the redshift bins in which the sample is divided, while the dashed lines show the threshold we used to define ``high mass`` and ``low mass`` groups (6$\times$10$^{13}$M$_{\odot}$).}
\label{fig1}
\end{center}
\end{figure}

The COSMOS field provides the largest catalog of X--ray selected groups obtained in a contiguous fields up to now. The catalog of COSMOS X--ray selected groups (status July 2010) contains 276 extended sources detected from a wavelet scale-wise reconstruction. The detection is performed on the co-added XMM-Newton and Chandra images, where point like sources have been subtracted from each dataset \citep{2009ApJ...704..564F,Leauthaud:2010p34}, and setting 4$\sigma$ as a threshold for the source detection.
A detailed description of the extraction of X--ray characteristics is given in \citet{Finoguenov:2007p475}.\\
The wealth of information available in the COSMOS data-base enables the optical identification of the groups both using photometric and spectroscopic data. 
Each group has been identified using a refined red-sequence technique as detailed in \citet{Finoguenov:2010p2012}. 
Furthermore, spectroscopic identification of groups has been achieved through the zCOSMOS-BRIGHT program \citep{Lilly:2009p1622}, targeted follow-up using IMACS/Magellan \citep{Finoguenov:2007p475} and FORS2/VLT (Finoguenov et al. in prep.), as well as through secondary targets on Keck runs by the COSMOS collaboration. \\
Following identification, the redshift of the individual groups is assigned on the basis of the available spectroscopy, or from the average photometric redshift of the red-sequence galaxies when less than two spectroscopic redshifts are available. 
The center of a galaxy group corresponds to the emission peak of the associated X-ray source. This X-ray center can be difficult to identify when the associated source is at the X-ray detection limit or a system is visually classified as a merger. In both cases a new center is assigned, which corresponds to the position of the most massive galaxy located near the X-ray center. 
A statistical treatment of the uncertainty in selecting  the groups' center can be found in \citet{George:2011p3935}. 
The robustness of the centring is evaluated through visual inspection and is expressed by a quality flag for each entry of the COSMOS X-ray group catalog. For the present analysis we have excluded groups having an uncertain optical counterpart or multiple ones (i.e. with a flag $>$ 3).\\
The X--ray detected groups span a large range of X--ray luminosities (5$\times$10$^{40}$--3$\times$10$^{43}$ erg/sec) and redshifts (0.08--1.9).
We limit the sample to 0.2$<$z$<$1.0, to ensure high quality photometric redshift \citep{Ilbert:2009p917} and a sufficient volume sampling. Also, we discard X--ray groups that fall outside the SUBARU area, and therefore have incomplete photometry (marked in red in Figure \ref{fig1}). 
After this selection we obtain a sample of 160 X--ray groups out of which 132 (82$\%$ of the sample) have at least three spectroscopic members within R$_{200}$ while 145 have at least two (90$\%$).   \\
The total masses of the X-ray groups are derived from the empirical L$_{X}$--M$_{200}$\footnote{M$_{200}$ is the mass enclosed in a circular region of radius R$_{200}$ within which the average density is 200 times the critical density of the universe at a given redshift.} relation determined in \citet{Leauthaud:2010p34} via weak lensing analysis. The resulting sample of X--ray detected groups ranges  between 1$\times$10$^{13}$ and 2$\times$10$^{14}$ \Msun in total mass with a median of 3.5$\times$10$^{13}$ \Msun. \\
We divide the group sample into 4 redshift bins between z=0.2--1.0, spanning 0.2 in redshift each (0.2--0.4, 0.4--0.6, 0.6--0.8, 0.8--1.0). \\
To study the behaviour of the galaxy stellar mass function as a function of the group mass, we divide the groups in two bins of M$_{200}$. We choose 6$\times$10$^{13}$ \Msun as the threshold  between low mass and high mass systems. This choice allows to maintain a similar median mass in the two mass bins and across the redshift range, maximizing the number of systems used. In this way, even if our sample is not mass complete for the lower mass groups, we can compare systems with on average similar properties at different redshifts. Nevertheless, at redshift 0.8--1.0 the median mass in the lower mass bin is a factor of two higher than in the lowest redshift bin, due to the decreased X--ray sensitivity to low mass systems. \\ 
Moreover due to the very low number of high mass systems in the redshift bins 0.4--0.6 and 0.6--0.8 (1 and 2, respectively), we choose  not to perform the analysis at these redshifts.\\
In Table \ref{tab_bins} we lists the characteristics of each groups subsample used in the following analysis; Figure \ref{fig1} shows the distribution of M$_{200}$ as a function of the redshift for the group sample and the division in subsamples.\\

\begin{table}
\centering
\caption{Characteristic of the Subsamples of COSMOS Groups}
\label{tab_bins}
\begin{tabular}{ccccc}
\hline

\hline
\multicolumn{1}{c}{Redshift}&\multicolumn{2}{c}{LOW MASS}&\multicolumn{2}{c}{HIGH MASS}\\
\hline
  & N &Log(M$_\mathrm{200}$)\tablefootmark{a}  [M$_{\odot}$]&N &Log(M$_\mathrm{200}$)$^{a}$  [M$_{\odot}$] \\
\hline
0.2$<$z$<$0.4&51 &2.1$\times$10$^{13}$&8&8.2$\times$10$^{13}$\\
0.4$<$z$<$0.6&20 &3.5$\times$10$^{13}$&3 &-\\
0.6$<$z$<$0.8&35  &3.5$\times$10$^{13}$&3&-\\
0.8$<$z$<$1.0&17 &4.2$\times$10$^{13}$&24&6.7$\times$10$^{13}$\\
\hline
\end{tabular}
\tablefoot{
\tablefoottext{a}{Median mass of the subsample.}
}

\end{table}%

\subsection{Galaxies in the COSMOS groups}\label{galaxies}
\begin{figure}
\begin{center}
\includegraphics[width=\columnwidth]{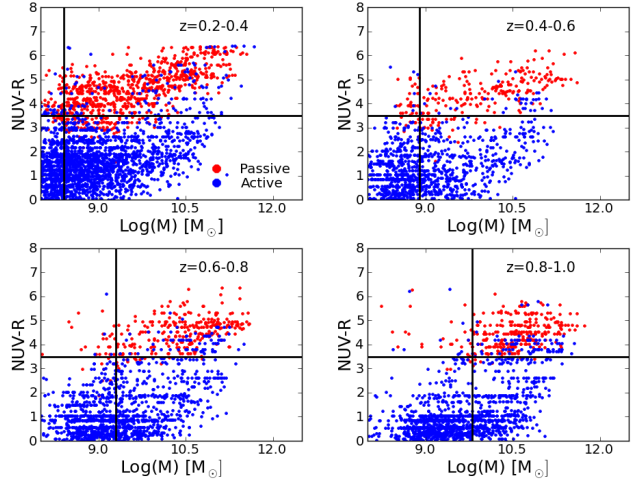}
\caption{NUV--R rest frame color (dust corrected) of galaxies within X--ray selected groups versus stellar mass. Red and blue symbols mark the classification as passive and star forming galaxies according to the spectrophotometric flag in the photometric catalog. Passive galaxies overlap with the associated sequence of red galaxies identified in rest frame NUV--R$>$3.5 and marked by the vertical line \citep{Ilbert:2009p917}. The solid horizontal line marks the completeness stellar mass for passive galaxies at each redshift.}
\label{NUVR}
\end{center}
\end{figure}

We use the COSMOS catalogue with photometric redshifts derived from 30 broad and medium bands described in \citet{Ilbert:2009p917} and \citet{Capak:2007p850} (version 1.8). 
We limit the galaxy selection to those brighter than i$^{+}_{AB}$=25, in order to ensure the accuracy of the photometric redshift to be within 0.03$\times$(1+z), as shown in Figure 9 in \citet{Ilbert:2009p917}. At this magnitude limit the detection completeness is $>$ 90$\%$ \citep{Capak:2007p850}. Furthermore we apply an additional infrared magnitude cut at K$<$24 to  limit the possible degeneracies in the photo-z and to ensure the reliability of the star/galaxy separation performed in the catalog by evaluation of the spectral energy distribution (SED) of each object.\\
The X--ray characteristics of galaxy groups provide R$_{200}$ as a scale radius to define individual systems. Candidate members are defined as all the galaxies within a projected distance equal to R$_{200}$ from the X-ray centroid of a group and within 0.02 × (1 + z) from its redshift (given in the X-ray catalog).%
To study the GSMF we use  the stellar mass of individual galaxies computed from their best-fit broad-band spectral energy distributions, as described in \citet{Ilbert:2010p420}, computed assuming a Chabrier initial mass function \citep{Chabrier:2003p4062}. The typical error on the stellar mass of a galaxy is 0.12 dex\footnote{The quoted error is the median of the errors on the individual stellar masses. Those are statistical errors computed once the model that best describes the galaxy SED is fixed. Larger systematics effects may be present when comparing mass estimated with a different set of models \citep{Longhetti:2009p712}.}, roughly half of that on the stellar mass estimated from the K-band absolute magnitude assuming a M/L ratio (see \citealt{Giodini:2009p923}).\\
It is worth stressing that both the stellar mass for individual galaxies and the initial mass function (IMF) used in this paper are different than those used in \citealt{Giodini:2009p923}, where masses where computed from  K-band photometry and the assumed IMF was that from \citet{Salpeter:1955p1869}. Differences between SED and K-band stellar masses are discussed in \citet{Ilbert:2010p420}, while changing between a Salpeter and a Chabrier IMF reduces the stellar masses of $\sim$0.25 dex.
Since we select a magnitude limited sample, we can ensure to observe all galaxies above a mass threshold (completeness stellar mass), which is redshift dependent.
To estimate the completeness mass we consider the galaxies in the faintest 20$\%$ of our sample and derive the stellar mass (M$_\mathrm{lim}$) they would have if their apparent magnitude was equal to the sample limiting magnitude (i.e. $i_\mathrm{AB}$=25). Then we define as completeness mass the value of the 95$\%$ percentile of the distribution in M$_\mathrm{lim}$  (the same method is applied in \citealt{2010A&A...523A..13P}): galaxies above this stellar mass limit define an 80$\%$ complete sample in stellar mass.
We calculate this at the upper limit of each of the redshift bins in which the groups' sample is divided, separately for star forming and passive galaxies (passive galaxies have a slightly higher completeness mass than star forming galaxies). The ensuing values represent the stellar mass completeness  as a function of redshift for our sample (Table \ref{tab_cmass}).\\
It is known that the distribution of galaxy properties is generally bimodal, being different for star-forming and passive galaxies \citep{2001AJ....122.1861S}. When studying galaxy evolution it is therefore very important to separate the two populations: for this purpose we use the spectroscopic types attributed to individual galaxies as a by--product of the photo-$z$ determination, on the basis of their best-fit broad-band spectral energy distributions (SEDs).
In particular, passive galaxies in the photometric catalog are  those which have as a best fit to the spectral energy distribution an early type galaxy template. In the COSMOS photometric catalog these galaxies have an SED type between 1 and 8 (for details on the templates see \citealt{Ilbert:2009p917,Polletta:2007p1833}). These SED types represent a  passive population consistent with an E/S0/Sa population selected morphologically \citep{Ilbert:2010p420}. As shown in Figure \ref{NUVR}, the galaxies in this category largely overlap with the associated sequence of red, passively evolving galaxies identified in rest frame NUV--R$>$3.5 (dust corrected), according to the classification of \citet{Ilbert:2009p917}. 
In particular, the spread of the passive galaxies population in  NUV--R at the lowest redshift bin is consistent with that found by \citet{Donahue:2010p2318} for brightest cluster galaxies in the REXCESS cluster sample. \\
Note that in Figure \ref{NUVR} the depletion of the red clump at low stellar masses at z$>$0.6  reflects the passive evolution of galaxies and not due to incompleteness: indeed, as shown in \citet{Juneau:2005p3499}, galaxies of M$_{stellar}<$10$^{10.8}$M$_{\odot}$ evolve from a bursting to quiescent star formation at z$\leq$1.\\
It is important to notice that our classification for a star forming galaxy is different from those based on spectroscopic information (e.g. [OII] or H$\alpha$ flux) or UV flux: the latter are sensitive only to very recent episodes of star formation (up to $\sim$10$^{8}$ years ago), while the SED contains also information from the rest frame optical emission which is 
sensitive to stellar populations with ages between 10$^{8}$--10$^{9}$ years. Having  this in mind, we can understand why some very red galaxies in Figure \ref{NUVR} are classified as star forming when considering the spectrophotometric classification.

\begin{table}
\centering
\caption{Completeness Galaxy Stellar Mass.}
\begin{tabular}{ccc}
\hline
\hline
Redshift & Log(M$_\mathrm{comp, passive}$) [M$_{\odot}$] & Log(M$_\mathrm{comp, SFG}$) [M$_{\odot}$]\\
\hline
0.2$<$z$<$0.4 & 8.6 &8.4 \\
0.4$<$z$<$0.6 & 9.1 &8.9 \\
0.6$<$z$<$0.8 & 9.6&9.3 \\
0.8$<$z$<$1.0 & 9.9&9.8 \\
\hline
\end{tabular}
\label{tab_cmass}
\end{table}%


\subsection{Galaxy stellar mass function of COSMOS groups}

\begin{figure}
\includegraphics[width=\columnwidth]{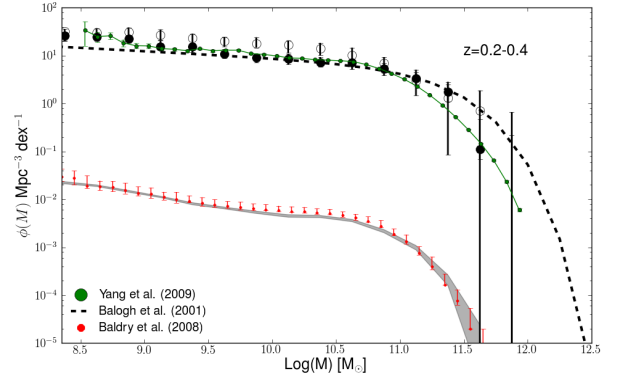}
\caption{Comparison of the galaxy stellar mass distribution for all galaxies in the COSMOS area (light gray) and X--ray detected low mass groups at z=0.2--0.4 (black circles). The dashed red line marks the GSMF obtained by Baldry et al. (2008) for the field. Green square show the GSMF of an optically selected groups sample from SDSS. The dashed black line marks the galaxy stellar mass function by Balogh et al. (2001) estimated for a sample of optically detected groups in the Las Campanas Redshift Survey \citep{Christlein:2000p4169} with velocity dispersion less the 400 km s$^{-1}$.}
\label{comparison}
\end{figure}

\begin{figure*}
\includegraphics[width=\textwidth]{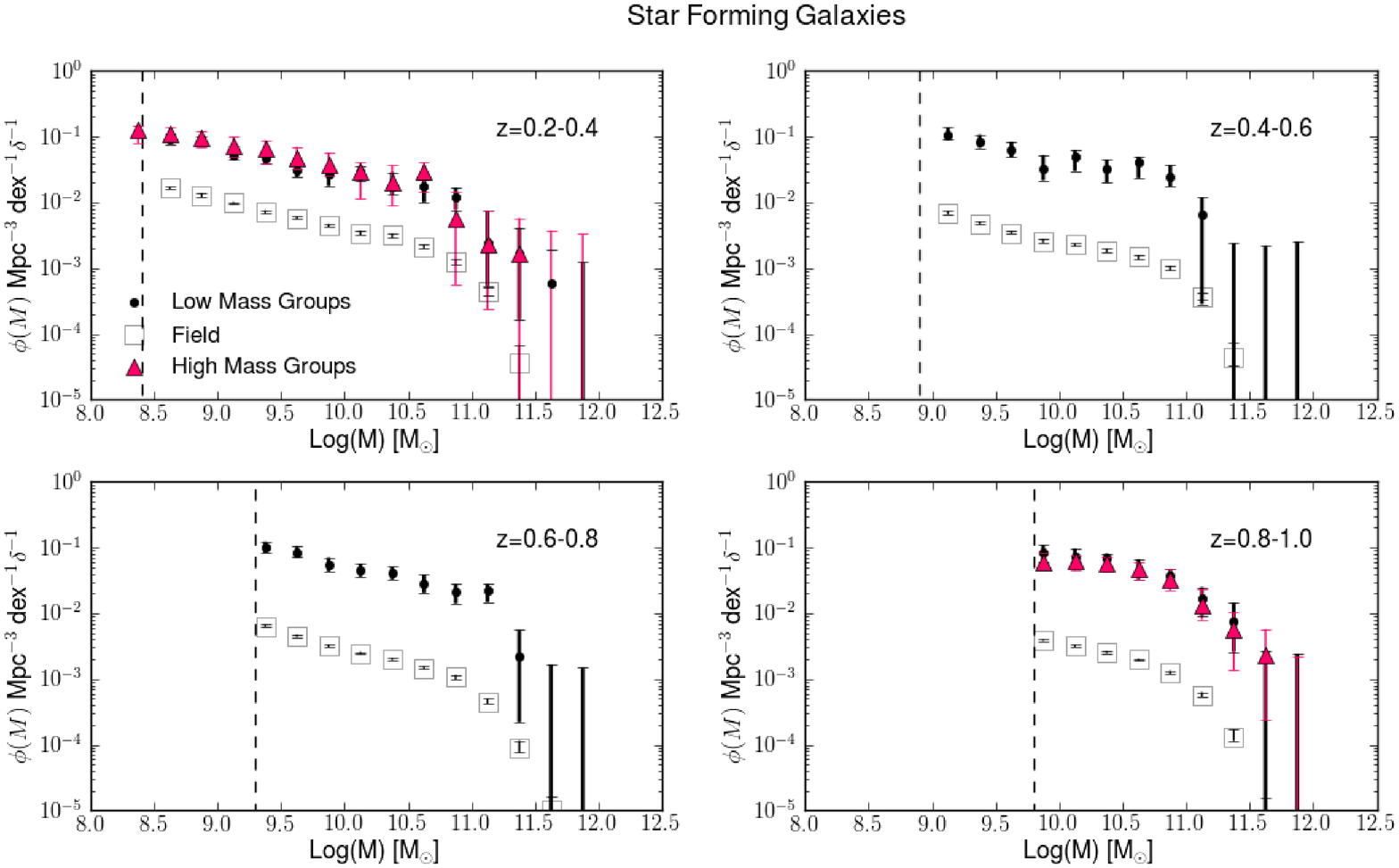}
\includegraphics[width=\textwidth]{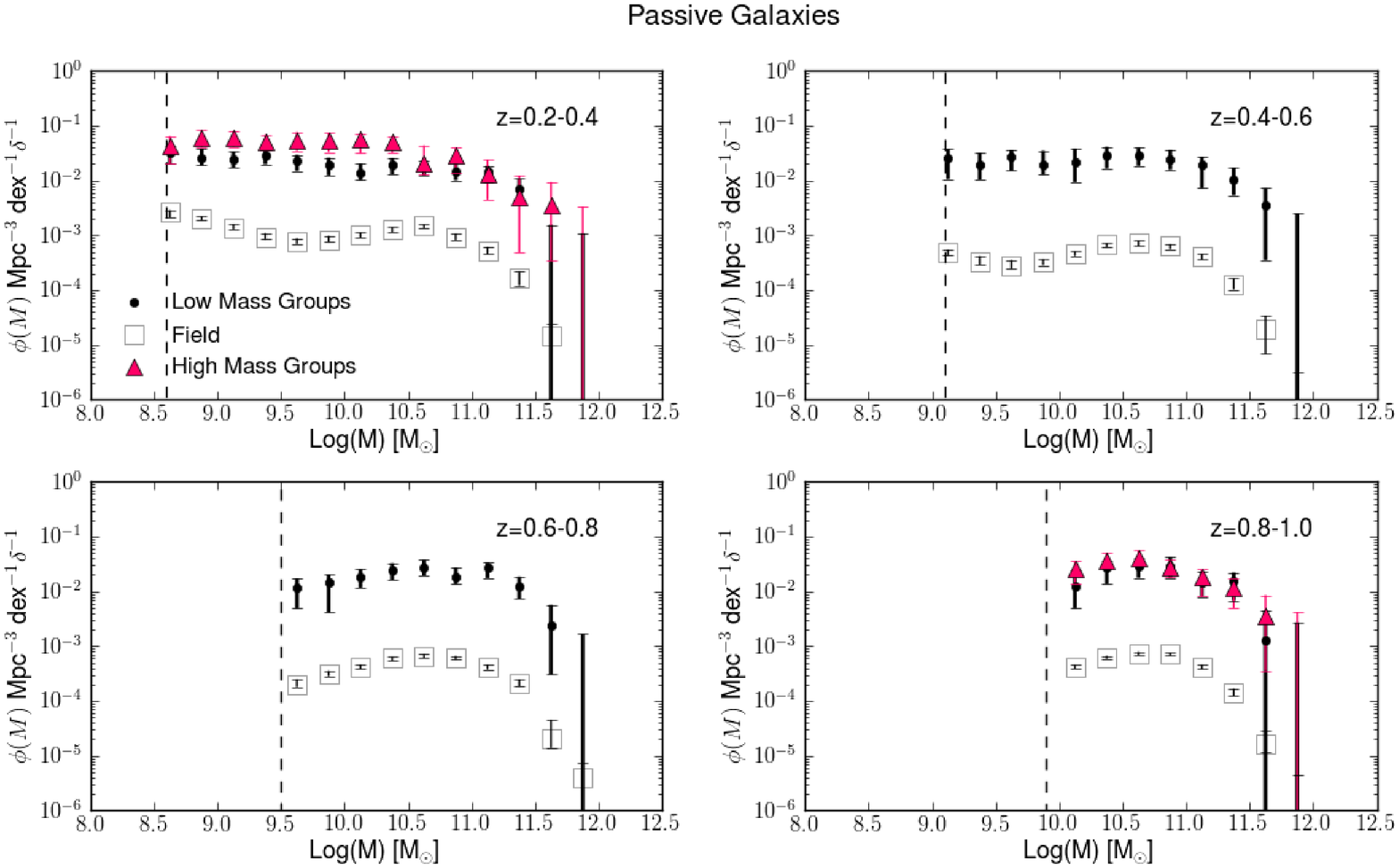}
\caption{\textit{Upper plot:} Galaxy stellar mass distributions of star forming galaxies in the field (squares),low mass groups (black points) and high mass groups (magenta triangles). Different panels show different redshift bins.  In each panel the vertical line marks the completeness stellar mass. The distributions are normalized to their respective overdensity  $\delta$ (over the critical one) at the redshift considered. The values of $\delta$ used are 200 for the groups while we assume an overdensity of 1  for the field. \textit{Lower plot:} Same as the upper panel but for passive galaxies. }
\label{gsmf2}
\end{figure*}

The distribution of galaxy stellar mass is obtained as follows.
For each bin of redshift and total mass the observed background subtracted mass distribution for the COSMOS X--ray selected groups can be expressed as:
\begin{equation}
\phi(M)=\frac{1}{V_{z}}\left[\left(\sum_{i}^{n}N(M)_{i}\right) -N_{b}(M)\right]
\end{equation}
where N indicates the member galaxies, N$_{b}$ is the contribution to the observed counts due to field galaxies, V$_{Z}$ is the volume sampled by X--ray groups  and n is the number of systems in each bin of redshift and total mass. The volume sampled by X--ray groups is computed as a sum of comoving spherical volumes with radius equal to R$_{200}$.\\
N(M) is obtained by direct counting of the member galaxies above the completeness stellar mass in bins of 0.25 dex in stellar mass. 
In order to obtain the composite stellar mass distribution in the group galaxies we statistically correct each stellar mass bin for the contribution of background galaxies by subtracting the background galaxy distribution.
The background distribution, N$_{b}$(M), consist of all the galaxies in the same redshift bin as the groups which are lying outside R${200}$ of any groups (hereafter called ``field galaxies"). This distribution is renormalized by a factor V$_{out}$/V$_{cylinder}$, with V$_{out}$ being the volume outside group's R$_{200}$ and V$_{cylinder}$ the volume where member galaxies are counted (see previous Section). We compute the background distribution from the whole survey field so that it is less affected by local fluctuations in the number density due to the large scale structure surrounding groups. However, works on the luminosity function based on SDSS data show that there is little difference between using a local and global background subtraction \citep{Goto:2003p3583, Popesso:2005p1365}. \\
It is worth noticing that our definition of field galaxies include those that do not reside in X-ray selected groups in our sample, but are not isolated and part of non-detected systems. In order to estimate possible contamination from surrounding large scale structure we repeat our analysis removing from the field galaxies' sample those within 1-4$\times$ R$_{200}$ of each groups (which mean considering only galaxies outside the turnaround radius of massive halos), and we do not find significant differences in the ensuing results.\\
The total uncertainties on $\phi(M)$  consist of the uncertainties on the stellar mass measurement and Poissonian errors.
The former are estimated via a Montecarlo simulation that redistributes the galaxies according to their 1$\sigma$ confidence interval on the stellar mass estimate (as given in the photometric catalog). The Poissonian uncertainty is the 1$\sigma$ variance for the low count regime computed as in \citet{Gehrels:1986p698}.\\
To test our determination of the GSMF in the groups and in the field, we compare it with other results at low redshift: in Figure \ref{comparison} we show the GSMF of all galaxies in low mass groups compared to those determined by \citet{Balogh:2001p86} and \citet{Yang:2009p343} using  2MASS and SDSS data respectively. The COSMOS field GSMF is compared with that of \citet{Baldry:2008p58}. \\
The GSMF we obtain for the field compares remarkably well with the independent determination by Baldry et al. (2008; red dashed line) on a spectroscopic sample, indicating that systematics on the stellar mass determination are under control, at least for the global distribution. The comparison with the groups' stellar mass distribution found by \citet{Yang:2009p343} is very encouraging, since the good agreement indicates that our group selection does not bias the galaxy distribution when compared to optically detected groups. We also find good agreement with the stellar mass distribution obtained by converting the K-band luminosity function found by \citet{Balogh:2001p86}. This suggests that our results can be easily compared also with those obtained in surveys where stellar masses for individual galaxies are not available. \\
Once tested that the global stellar mass distributions are robust when compared with previous studies, we proceed dividing the galaxy population between star forming and passive objects.
In Figure \ref{gsmf2} we show the composite, background subtracted stellar mass distributions of star-forming and passive galaxies in the COSMOS X--ray selected groups. The contribution of low and high mass groups is considered separately. Given the small number of systems, the stellar mass distribution of high mass groups at redshifts 0.4--0.6 and 0.6--0.8 cannot be robustly determined and is not used for the analysis.\\
We also show as a comparison the distribution of galaxy stellar mass in the field, defined as all coeval galaxies outside bound X--ray emitting structures. To enable the comparison in Figure \ref{gsmf2}, we normalize the distributions to their respective overdensity  $\delta$ (over the critical one) at the redshift considered. The values of $\delta$ used are 200 for the groups and 1  for the field.\\
In Appendix \ref{table_gsmf} we list the galaxy stellar mass distributions for low and high mass groups in COSMOS at different redshifts.

\section{Parametrization of the composite GSMF}\label{results}

\subsection{Star Forming Galaxies}\label{active}
\begin{figure*}
\begin{minipage}[b]{\textwidth}
\centering
\includegraphics[width=\textwidth]{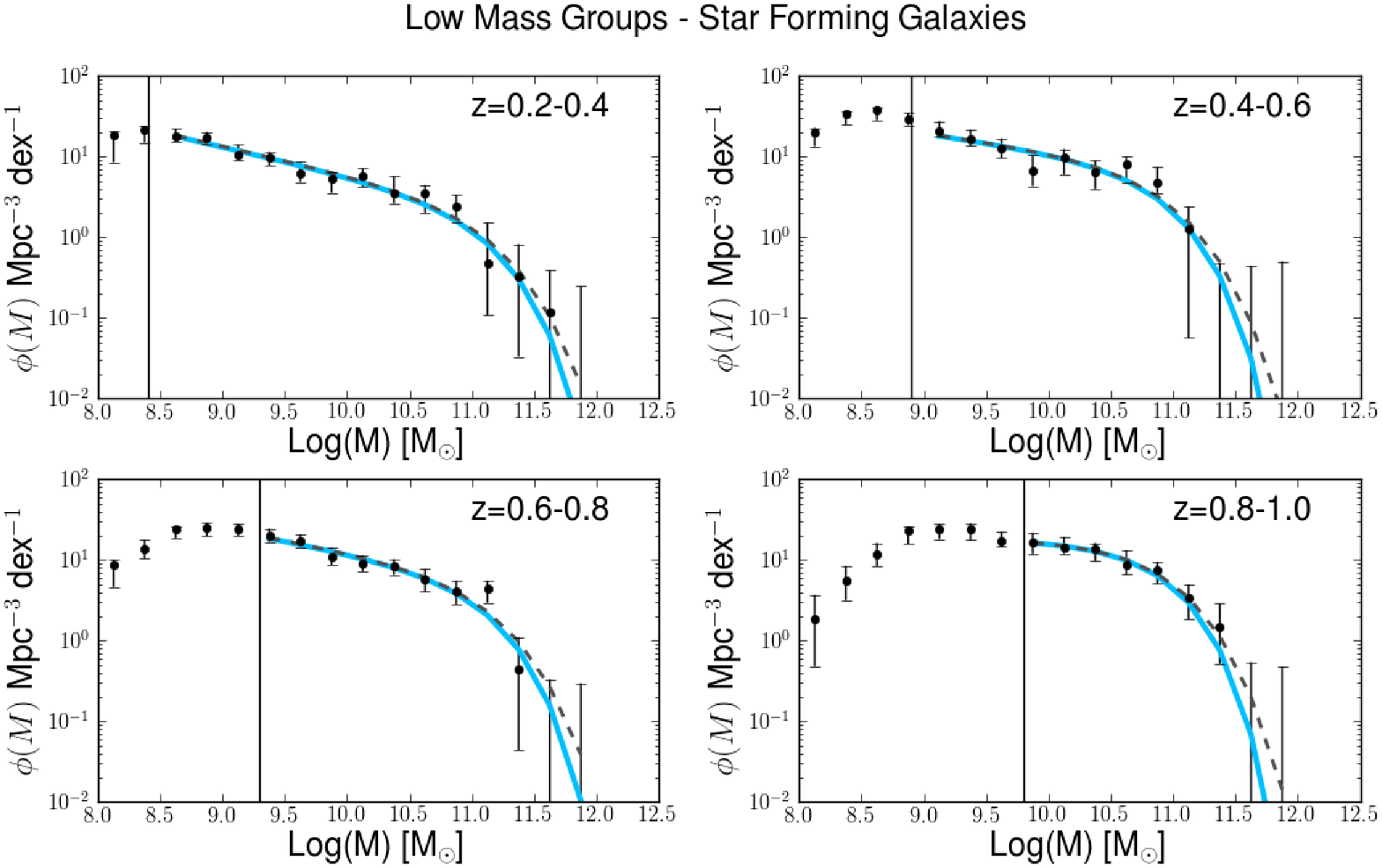}
\caption{Schechter function fit to the GSMF for star forming galaxies in low mass groups. The dashed line marks the convolved Schechter function and  the solid one the ``true'' Schechter function (all fitting parameters free). The vertical line marks the completeness stellar mass.}
\label{act_fit_low}
\end{minipage}  
\begin{minipage}[b]{\textwidth}
\centering
\includegraphics[width=\textwidth]{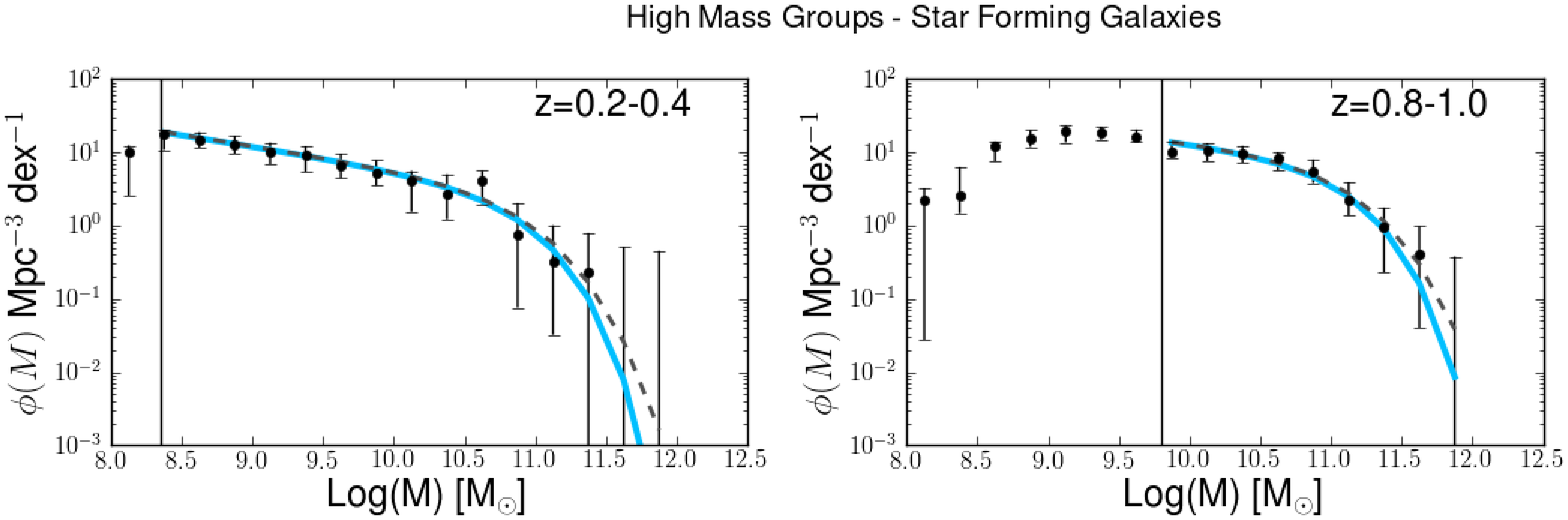}
\caption{Schechter function fit to the GSMF for star forming galaxies in high mass groups. The dashed line marks the convolved Schechter function and  the solid one the ``true'' Schechter function (all fitting parameters free). The vertical line marks the completeness stellar mass. }
\label{act_fit_high}
\end{minipage}
\end{figure*}

\begin{figure}
\includegraphics[width=\columnwidth]{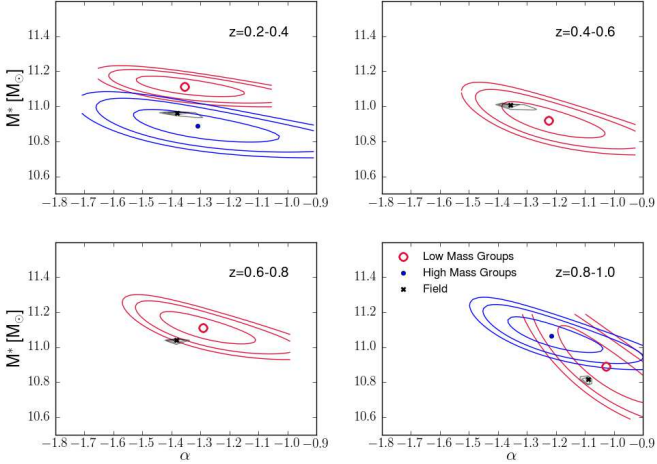}
\caption{Confidence ellipses on the Schechter parameters estimated for the mass function of star forming galaxies in the field (black), low mass groups (red) and high mass groups (blue).  Contours correspond to $-2\mathrm{Log}(\ell)$ differences above the maximum of 2.30,6.17 and 9.21 representing confidence level of  $\alpha$ equal to 68.3, 95.4 and 99$\%$  respectively.}
\label{banana}
\end{figure}

The binned distribution of stellar mass for star forming galaxies can be described by a Schechter function with slope $\alpha$ and characteristic mass \ms:
\begin{equation}
\phi_{s}(\mathrm{Log(M)})=\phi^{*}\left(\frac{\mathrm{Log(M)}}{\mathrm{Log(M^*)}}\right)^{\alpha +1}e^{-\frac{\mathrm{Log(M)}}{\mathrm{Log(M^*)}}}
\end{equation}
where M is the stellar mass.\\
We perform Montecarlo simulations of the galaxy distribution to quantify the effect of uncertainties on the shape of the observed galaxy stellar mass function. 
The observed distribution is modified by the effect of uncertainties on the stellar mass, that convolves the true mass distribution with the stellar mass error distribution. 
We can take this effect into account by modelling the stellar mass error distribution. The  distribution of stellar mass uncertainties can be described at the first order as a Gaussian in log-space with rms $\sigma_{Log(M)}$ equal to 0.12 dex, which is the typical error on the stellar mass measurement (see Section \ref{galaxies}). Montecarlo simulations of the galaxy distribution confirmed that the true distribution is correctly recovered by using a single gaussian convolution, even if the error distribution may be more complicated.\\
Therefore the observed galaxy stellar mass function is then given by:
\begin{equation}
\phi(\mathrm{Log(M)})=\frac{1}{\sqrt{2\pi}\sigma_{\mathrm{Log(M)}}}\int{\phi_s}\,e^{\frac{\mathrm{-(Log(M^{'})-Log(M))^2}}{\mathrm{2\sigma_{Log(M)}^{2}}}}\,d\mathrm{Log(M^{'}})
\end{equation}
In galaxy groups the observed stellar mass distributions are the sum of the background and the true Schechter function convolved with the distribution of uncertainties.
By fitting this convolved function to the observations, the Schechter function $\phi_{s}$ thus determined describes the ``true'' mass function which would be measured in absence of uncertainties on the stellar mass.\\
Due to the low number of counts in some stellar mass bins (at high stellar mass), $\chi^2$ minimization is not an appropriate technique and therefore we use a maximum likelihood fitting method.
We maximize the logarithmic likelihood (Log($\ell$)) that the model may describe our data with respect to the Schechter parameters; 1$\sigma$ confidence levels on the best fitting parameters are obtained by identifying the interval at which $-2\mathrm{Log}(\ell)$ is lower by 1 than at its maximum. 
We verified through Montecarlo simulations that our method could recover the correct Schechter function, as well as the size of the parameter uncertainties.\\
The Schechter parameters obtained for the most likely function are listed in Table \ref{tab_param}, together with the 1$\sigma$ confidence intervals. 
In Figure \ref{act_fit_low} and \ref{act_fit_high} we show the best fit Schechter functions (in cyan) for star forming galaxies in low mass and high mass groups respectively, plotted over the observed background subtracted stellar mass distribution.\\
A visual inspection of Figure \ref{gsmf2} (upper panel) suggests that the distributions at low stellar mass seem remarkably similar in groups and the field at all redshifts. We can confirm this in a quantitative way by comparing the values of $\alpha$ found in groups and the field. The slope is compatible with being the same in all environments and it does not show a significant evolution over the whole redshift range (but for the last redshift bin, where the derived slope may be artificially flat as discussed in the next section).\\ 
This similarity is not confirmed when looking at the high stellar mass part of the distribution. At low redshifts the characteristic mass of low mass and high mass groups is offset, with the one of low mass groups being higher. When considering the single-parameter uncertainties on M$^{*}$ (listed in Table \ref{tab_param} ), the discrepancy is not statistically significant ($\sim$1$\sigma$). Nevertheless it is interesting to consider the effect of the correlation with the slope $\alpha$ as shown in Figure \ref{banana}, where we draw the iso-likelihood contour for the  combination of best fitting $\alpha$ and M$^{*}$, corresponding to  68.3, 95.4 and 99$\%$ confidence levels (corresponding to 1, 2, 2.5 sigma). \\ As can be seen from the plot, for any common value of $\alpha$ between the high and low mass groups contour region, the M$^{*}$ is different at more than 2$\sigma$. Therefore if $\alpha$ is the same in high and low mass groups, the significance of a discrepancy in the characteristic mass is enhanced, which suggest the bulk of star forming galaxies being more massive in low mass groups.\\
Furthermore, the values of characteristic mass we find for star forming galaxies in the field and in groups are consistent with those estimated in \citet{Bolzonella:2009p425}  for low density (D1) and high density (D4) environments, respectively. A hint of the offset we observe between M$^{*}$ in differently dense environment could be seen also using the \citet{Bolzonella:2009p425} values. However the large uncertainties on their estimates strongly affect any conclusion in this respect. \\ 
We also note that the characteristic mass of star forming galaxies in all the environments is remarkably stable across the redshift range. This finding confirms aand extend to the regime of groups that by \citet{Ilbert:2010p420}, who found the same redshift independence in the global galaxy stellar mass distribution.\\

\begin{figure}
\includegraphics[width=\columnwidth]{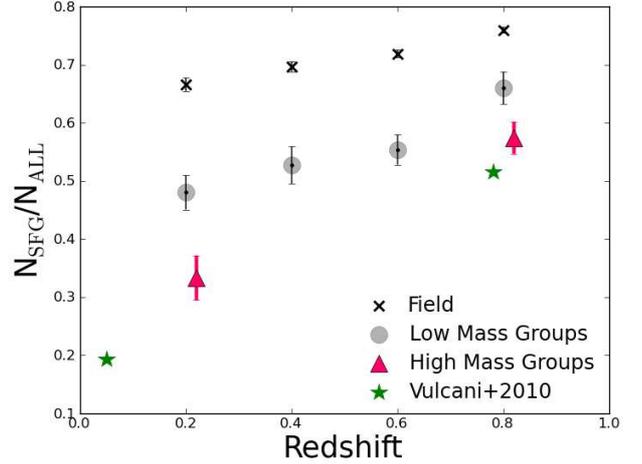}
\caption{Fraction of star forming galaxies above Log(M/M$_{\odot}$)=9.8 in the field (black crosses), low mass (grey large circles), high mass groups (magenta triangles).  The green stars mark the fraction of late type galaxies observed in massive clusters by \citet{Vulcani:2010p919}}
\label{frac_act}
\end{figure}

On the other hand it is known that the fraction of star forming galaxies decreases with time in massive clusters (Butcher-Oemler effect; \citealt{1978ApJ...219...18B}) and that denser environments are dominated by non star forming galaxies. \\ Figure \ref{frac_act} shows the fraction of star forming galaxies with Log($\frac{M_\mathrm{stellar}}{M_{\odot}}$)$>$9.8 in the different environments and as a function of redshift. In this plot we can appreciate how the star forming galaxies' fraction decreases towards lower redshifts in all the environments, but at different paces. The fraction of star forming galaxies in the field evolves slowly with time, confirming the slow evolution in the field, where environmental processes are less important. 
Within low and high mass groups, instead, the decrease in the contribution of star forming galaxies as a function of redshift is more noticeable and stronger in the most massive systems. Interestingly at redshifts 0.2--0.4 low mass groups exhibit 50$\%$ more star forming galaxies than high mass groups. 
We compare our result with that of \citet{Vulcani:2010p919} who use sample of massive systems (with velocity dispersion larger than 500 km/s, which translates roughly in M$_{200}>$10$^{14}$ \Msun) at redshift $\sim$0.1 and $\sim$0.8.  
These clusters exhibit an even lower star-forming galaxy fraction at low redshift, when compared with the COSMOS groups. This finding points towards galaxy groups being  an intermediate environment in terms of star forming galaxy content, when compared to lower and higher mass systems, and it indicates an extension of the Butcher-Oemler effect to galaxy groups. It is also interesting to note that the differences among the various environments increase with time, indicating a faster evolution of the fraction of star forming galaxies in more massive groups.

\subsection{Considerations}\label{considerations}
When fitting a Schechter function to the galaxy stellar mass distribution, we consider only the stellar mass bins not affected by incompleteness, in order to ensure a robust estimation of the parameters. If the model adequately represents the data, restricting the range where the fit is performed should not affect the final result (except for increasing the uncertainties on the parameters). 
However, when using at redshift 0.2--0.4 the same limit in stellar mass  as at high redshift, the resulting best fit slope is flatter (near to -1.0) and the associated error increases by a factor of two. Therefore, the slope estimated in the previous section at z=0.2--0.4 ($\sim$-1.4) is still within 1.5 $\sigma$ the newly estimated one, while the same is not true at  high redshift where the steep slope found at low redshift is rejected at more than 3 sigma significance. 
As a further test, we perform a fit of 50 Montecarlo realizations of the observed GSMF of low mass groups at z=0.2--0.4, using logM=8.35 and 10.0 \Msun as limiting masses. While the former leads to a slope steeper than -1.3 as best fit solution in the majority of the cases, the same happens only in 11 out of 50 cases when using the highest limiting mass. Furthermore the means of the two distributions of best fitting slopes with high and low limiting mass differ at $\sim$2 sigma significance.\\
The previous test tells us that probably a single Schechter function is not the most adequate model to describe the stellar mass distribution of star forming galaxies, and that more information is contained in the low mass part of the distribution. However, it is possible to pinpoint the steep slope of the function only with deep data (e.g. log(M/\Msun)$<$10.0), and this may explain why other works on galaxy groups performed with shallower data find flatter slopes than ours (e.g. Yang et al 2009).
However, since the aim of this paper is a first characterization of the GSMF of galaxy groups to be compared with other observations and galaxy evolution models which adopt a single Schechter description, we also assume this prescription and leave a more detailed analysis of the shape of the low mass GSMF to a future work.\\
The test above suggests that at high redshift, the slope may be artificially flat, due to the lack of deeper data. The data points below the completeness mass can potentially
identify an artificially flat slope, as they represents lower limits. For high-mass groups at redshift 0.8-1.0, the first two points below the completeness mass are
higher than the prediction from the best-fit model with a flat slope. We therefore re-fit the slope including these two data points, and find a steeper slope ($\sim$1.2) compatible within 1$\sigma$ with that found at low redshift. In the following we use this result instead of the artificially flat fit above the completeness mass, noting that if we would have deeper data, the slope would even be somewhat steeper.

\renewcommand{\arraystretch}{1.5}
\begin{table*}[!h]
\caption[width=\textwidth]{\large Schechter Best Fit Parameters to the GSMF of Star Forming Galaxies}
\label{tab_param}
\setlength\tablinesep{3pt}
\begin{center}
\begin{tabular}{c c c c c c c}
\hline
\hline
   z & V [Mpc$^{3}$] &$\phi$ &  log(\ms h$^{-2}$)&$\alpha$ & $\phi_{\alpha=-1.4}$\tablefootmark{a} &  log(\ms$_{\alpha=-1.4}$ h$^{-2}$)\tablefootmark{a}\\
\hline
\multicolumn{7}{c}{LOW MASS GROUPS}\\
\hline
0.2-0.4& 34.28&20.20$^{6.00}_{6.00}$&10.82$^{0.14}_{0.12}$&-1.35$^{0.04}_{0.04}$&14.96$^{1.80}_{1.72}$&10.92$^{0.11}_{0.10}$\\
0.4-0.6& 16.81&30.28$^{10.00}_{8.00}$&10.62$^{0.14}_{0.12}$&-1.23$^{0.08}_{0.08}$&13.49$^{1.88}_{1.76}$&10.89$^{0.12}_{0.10}$\\
0.6-0.8& 25.20&37.13$^{12.00}_{10.00}$&10.81$^{0.12}_{0.12}$&-1.29$^{0.08}_{0.08}$&22.46$^{2.60}_{2.48}$&10.97$^{0.09}_{0.09}$\\
0.8-1.0& 15.63&65.86$^{24.00}_{20.00}$&10.59$^{0.14}_{0.12}$&-1.03$^{0.16}_{0.16}$&23.18$^{3.48}_{3.20}$&10.92$^{0.10}_{0.09}$\\

\hline
\multicolumn{7}{c}{HIGH MASS GROUPS}\\
\hline
0.2-0.4& 11.30&12.90$^{6.00}_{6.00}$&10.59$^{0.18}_{0.16}$&-1.31$^{0.08}_{0.08}$&7.55$^{1.32}_{1.24}$&10.77$^{0.16}_{0.16}$\\
0.8-1.0& 16.80&39.70$^{12.00}_{10.00}$&10.77$^{0.12}_{0.12}$&-1.21$^{0.08}_{0.08}$&19.57$^{3.16}_{2.76}$&10.99$^{0.10}_{0.12}$\\

\hline
\multicolumn{7}{c}{FIELD}\\
\hline
0.2-0.4&0.56 $\times$ 10$^{6}$&273.02$^{20.00}_{20.00}$&10.67$^{0.04}_{0.04}$&-1.38$^{0.02}_{0.02}$&242.38$^{5.00}_{5.00}$&10.71$^{0.00}_{0.05}$\\
0.4-0.6&1.25 $\times$ 10$^{6}$&397.45$^{30.00}_{28.00}$&10.71$^{0.04}_{0.04}$&-1.36$^{0.02}_{0.02}$&318.60$^{10.00}_{10.00}$&10.79$^{0.10}_{0.05}$\\
0.6-0.8&1.97 $\times$ 10$^{6}$&656.20$^{46.00}_{44.00}$&10.74$^{0.04}_{0.04}$&-1.38$^{0.02}_{0.02}$&608.27$^{15.00}_{15.00}$&10.77$^{0.10}_{0.05}$\\
0.8-1.0&2.62 $\times$ 10$^{6}$&2252.97$^{136.00}_{132.00}$&10.52$^{0.04}_{0.04}$&-1.09$^{0.04}_{0.04}$&965.19$^{25.00}_{25.00}$&10.79$^{0.15}_{0.05}$\\

\hline
\hline
\end{tabular} 

\tablefoot{
\tablefoottext{a}{Obtained fixing the slope $\alpha$ to -1.4}
}
\end{center}
\end{table*}

\subsection{Passive Galaxies}
\begin{figure*}
\begin{minipage}[b]{\textwidth}
\centering
\includegraphics[width=\textwidth]{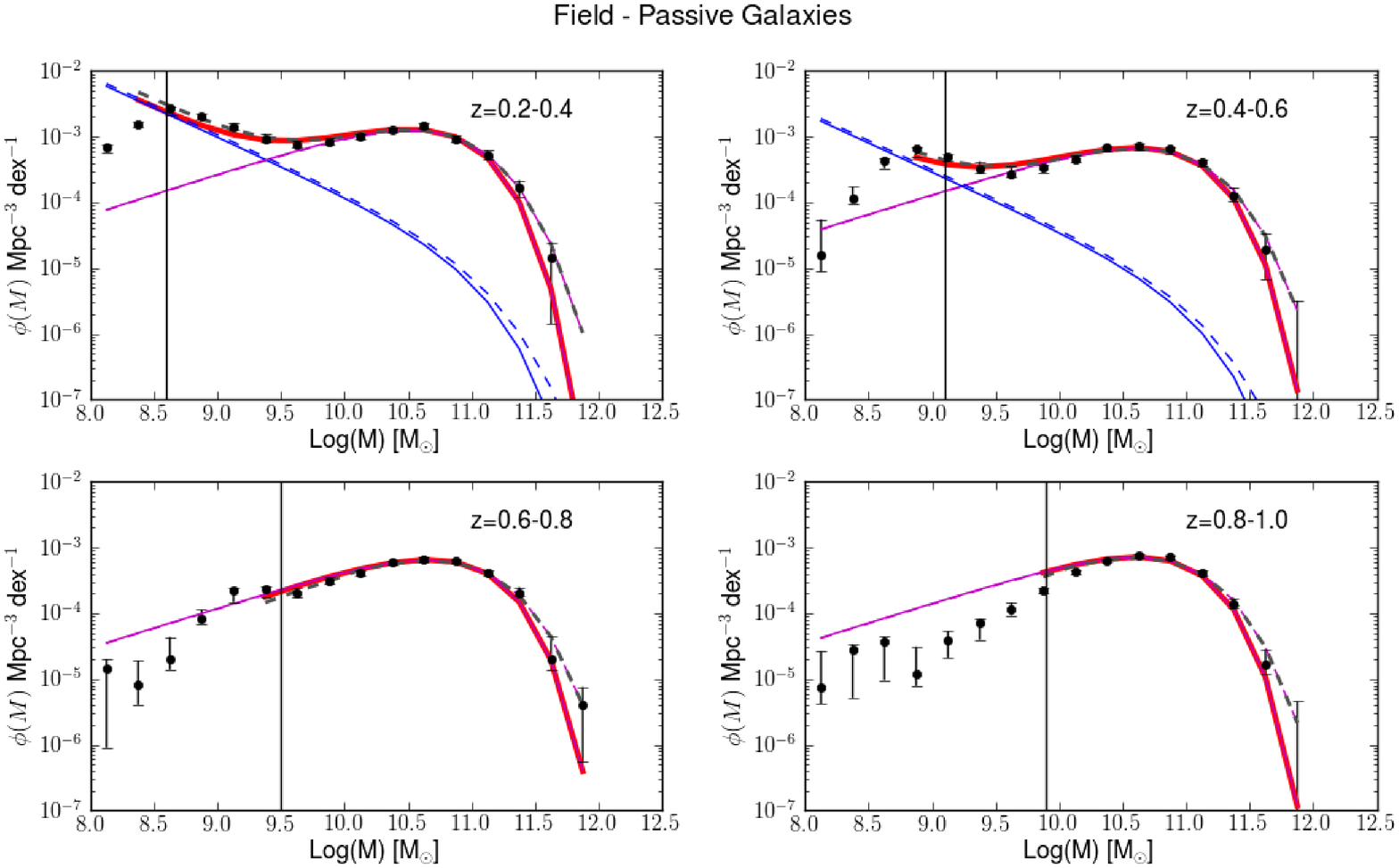}
\includegraphics[width=\textwidth]{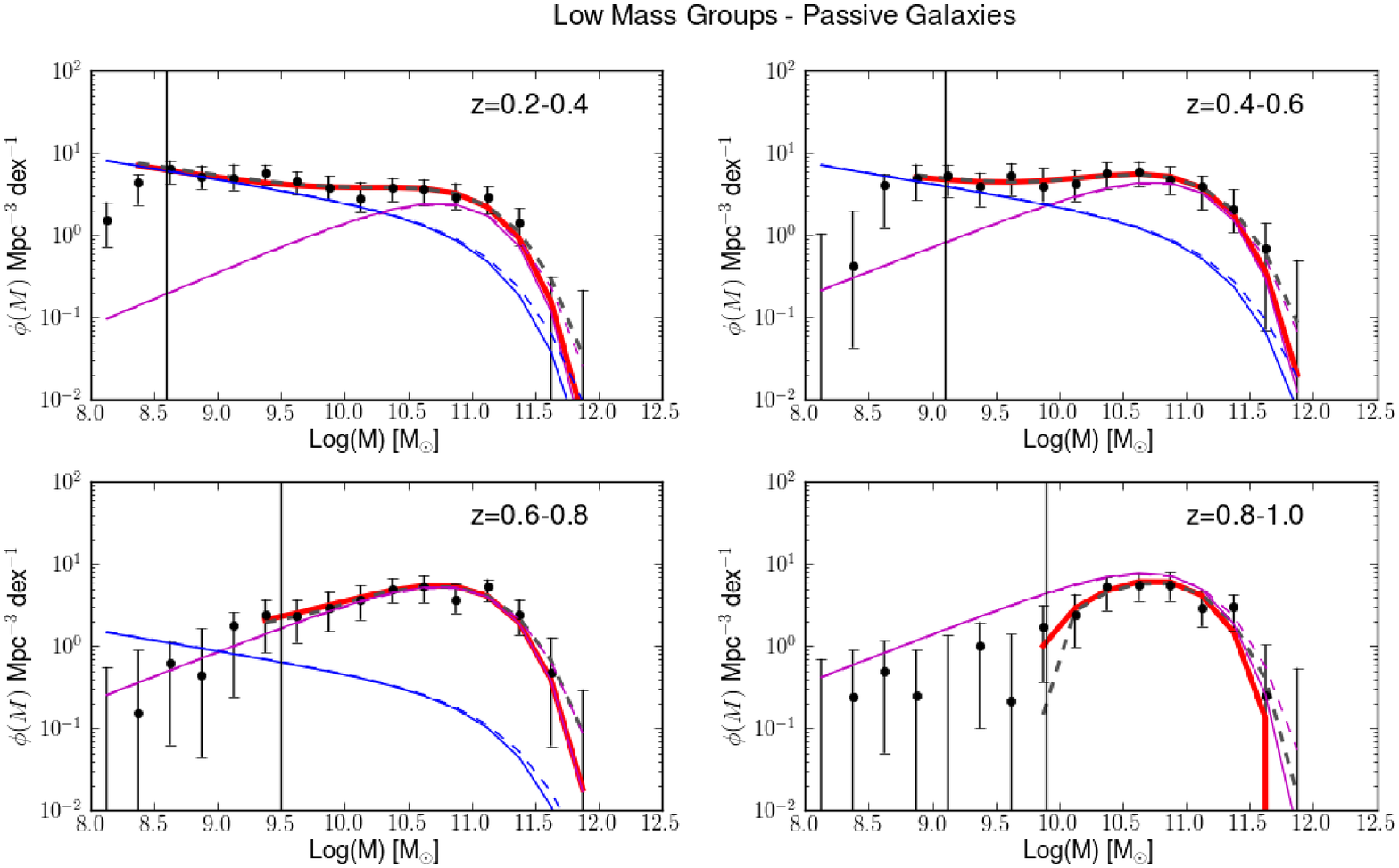}
\caption{\textit{Upper Panel}:Schechter function fit to the GSMF for passive galaxies in the field. The dashed line marks the convolved Schechter function, the solid one the ``true'' Schechter function. The magenta and blue lines represents the two components of the double Schechter function (in red). The solid vertical line marks the completeness stellar mass. \textit{Lower Panel:} the same as for the upper panel but for low mass groups.}
\label{etg_fit_field}
\end{minipage}
\end{figure*}

\begin{figure*}
\begin{minipage}[b]{\textwidth}
\centering
\includegraphics[width=\textwidth]{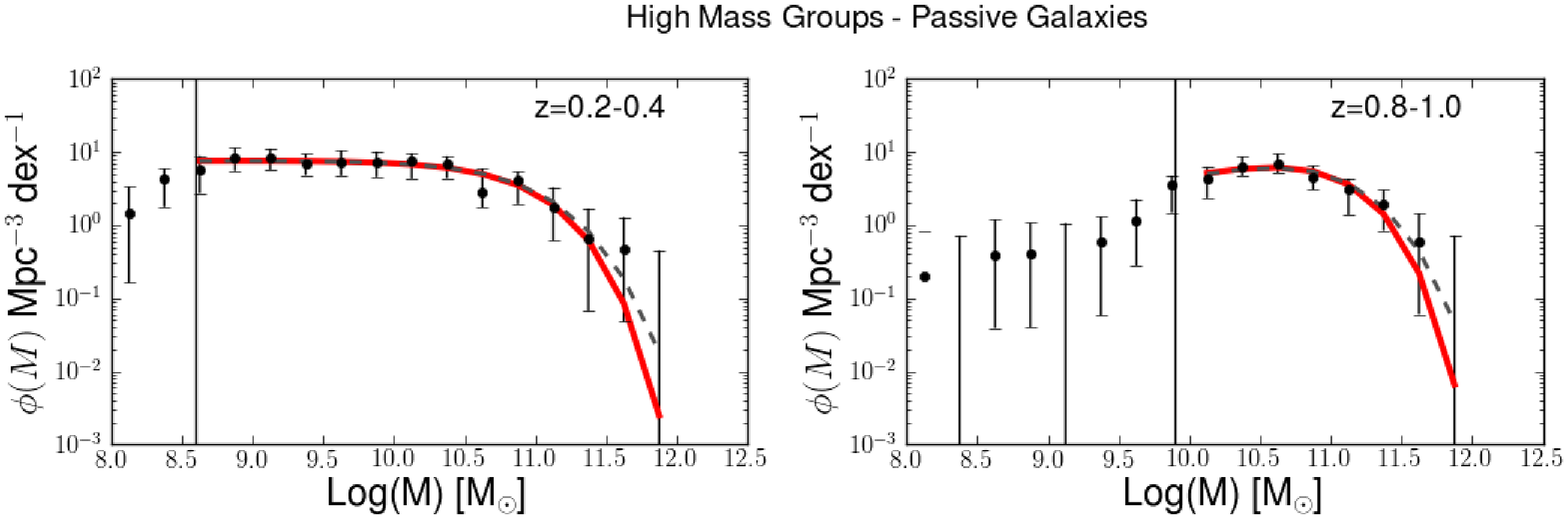}
\caption{Schechter function fit to the GSMF for passive galaxies in high mass groups. The dashed line represents the convolved Schechter function, the solid one the ``true'' Schechter function. The GSMF of massive groups is  adequately described by a single Schechter function (see text). The solid vertical line marks the completeness stellar mass.}
\label{etg_fit_high}
\end{minipage}  
\end{figure*}

As can be seen in Fig. \ref{gsmf2}, the observed stellar mass distribution of passive galaxies is characterized by a different shape than that of star forming galaxies in both subsamples of groups and the field.  
In groups the distributions flatten at low masses. Conversely, in the field the low mass distribution exhibits a ``dip" around log(M/\Msun)$\sim$9.5-10.0.\\ This behaviour can be hardly described by a single Schechter function: previous works suggested that this more complicated shape is produced by two different populations of galaxies, each with a GSMF described by a single Schechter function. Following this prescription we parametrize the galaxy stellar mass function with a sum of two Schechter functions where $\phi_2$ accounts for the steep rising slope towards low masses (secondary component) and $\phi_1$ for the flatter slope at high masses (primary component). \\
If the galaxy stellar mass function stems from two classes of galaxies, it is reasonable to assume that at least one of them has something in common with the distribution of star forming galaxies, especially if the process that has quenched the star formation is fast enough to prevent their further growth in stellar mass via star formation. \\
It is also reasonable to assume that this would happen more likely for galaxies of low mass, that are more strongly affected by environmental effects. Therefore, assuming that low mass quenched galaxies stem directly from the distribution of star forming galaxies, we set the M$^*$ of the low mass component (M$^{*}_{2}$) to that found for star forming galaxies. \\
On the other hand, massive galaxies are generally segregated towards the center of the potential well of a group because of dynamical friction, where merging episodes are more likely to occur. This may produce some change in the characteristic mass of the primary component from that of star forming galaxies, therefore we leave this parameter as free. \\
It is also true that M$^*$ for star forming galaxies is remarkably stable across the cosmic time. Thus we may assume that star formation and  its shut off are acting at the same pace for massive galaxies  following \citet{Peng:2010p568}.
Translating this assumption into requirements on the fitting parameters, the $\alpha_1$ can be fixed to 
\begin{equation}
\alpha_{1}=\alpha_{SFG}+(1+\beta)
\end{equation}
where $\beta$ is the slope of the specific star formation rate (sSFR)--stellar mass relation \citep{Peng:2010p568}. In first approximation $\beta$ can be set to zero, since the dependence of the sSFR on the stellar mass is found to be very weak \citep{2007A&A...468...33E, Noeske:2007p1019}. However, given that at high redshift $\alpha_{SFG}$  can be eventually affected by systematics discussed in Section \ref{considerations}, we also consider the case where the value  of $\alpha_{1}$ is fixed to -0.4 at z$>$0.4 . \\
Furthermore, since at redshift larger than 0.4 the high completeness mass prevents us from constraining the slope of the secondary component, we fixed it to that found at low redshift.\\
In Figures \ref{etg_fit_field} and \ref{etg_fit_high} we show the resulting best fit Schechter functions over--plotted to the background subtracted observed stellar mass distribution for low and high mass groups respectively. \\
In  Table \ref{tab_param_etg} we list the best fitting Schechter parameters. In some cases at high redshift we are unable to constrain the secondary component, and the formal best fit $\phi_{2}$ results in a negative value. In these cases we did not fit the second component and we set $\phi_{2}$=0.\\

We find evidence for a steeply rising slope at low stellar masses in the mass function of passive galaxies in the field.
To ensure the robustness of the measure against a contamination from star forming galaxies, we repeat the slope estimation with the additional condition of SFR$<$10$^{-2}$ \Msun y$^{-1}$ and NUV-R color larger than 3.5. The latter is the same condition applied in \citet{Ilbert:2009p917} to select quiescent galaxies. In both cases the steepness of the slope is confirmed.\\
 We also test against contamination from catastrophic errors in the photometric redshift determination of high redshift galaxies. Such galaxies, being faint, will be assigned a low mass and may contribute to the lower part of the galaxy stellar mass distribution. 
 We scale the distribution in $i$ magnitude of the galaxies with z$>$1 by an upper limit of 20$\%$ catastrophic failures, as estimated by \citet{Ilbert:2009p917} at $i>$23, and we find that the contamination becomes important only at $i>$26, well beyond our limiting magnitude. Therefore contamination from misidentified high redshift galaxies can be ruled out as an explanation for the steepness of the observed slope.\\ 
 We also test against the contamination from the outskirts of the large scale structure by repeating the analysis with a more conservative selection of the background, removing galaxies within 3$\times$R$_{200}$ from the center of a group. Also in this case the GSMF show a steep slope at low mass, confirming the robustness of our finding as a feature of the field.\\
At low redshift, the secondary slope in the field appears to be significantly steeper than that found for star forming galaxies ($\sim$-1.8 when compared to $\sim$-1.4, with a difference at the 7$\sigma$ level). This indicates that the quenching of low mass galaxies may be more complicated than that predicted by simple models of galaxy evolution \citep{Peng:2010p568}. Indeed, if the observed stellar mass distribution of star forming galaxies is the same as that of the progenitors of the quenched galaxies, our finding suggests that during the process of quenching this distribution is not conserved. Unfortunately we cannot well constrain the slope at higher redshift due to incompleteness, but we can fit the points also below the completeness threshold to obtain a lower limit to the slope. The estimated lower limits indicate that a rising slope at low stellar masses exists at least since redshift 0.6. We do not draw any conclusion on the highest redshift bin since the points at Log(M$_\mathrm{stellar}<10$ are likely affected by strong incompleteness. Indeed we observe a strong evolution in the fraction of passive galaxies with stellar mass Log(M$_\mathrm{stellar}<10$ at these redshifts (see Figure\ref{NUVR}) which is not found in studies on the global population of galaxies \citep[e.g.]{2006ApJ...651..120B}. \\
Interestingly, such a steep slope is not found in groups, where $\alpha$ tends to be less negative in higher mass groups. As a test we perform a fit of the groups data-points with a double Schechter function having a slope as steep as the field. This functional form does not adequately describe our data: it underestimates the number density of galaxies at intermediate mass and strongly overestimate that of dwarf galaxies within structures. Therefore we conclude that the distribution of the field and groups' GSMF cannot be described by a function with the same value of $\alpha_{2}$.\\
When examining the change in the shape of the GSMF it is clear the growing influence of the environmental effects switching from field to low and high mass groups. In particular the environmental component rises so strongly in high mass groups that it is impossible to obtain a robust fit for the primary component. We therefore choose to fit high mass groups as a single Schechter function.\\
Finally our M$^{*}$ values for the high-mass component are 0.1--0.2 dex larger M$^{*}$ for groups compared to the field at all redshifts, corresponding to a growth in mass by $\sim$40$\%$. Merging after the shut off of star formation, thus called ``dry'' merging \citep[e.g.]{vanDokkum:2005p4410}, can contribute to the growth of the mass of passive galaxies in groups. Our estimate is in agreement with estimates at redshift zero, according to which dry mergers should not increase the mass of passive massive galaxies of more than $\sim$45$\%$ (Nipoti et al. 2009).   \\
It is worth stressing that a double Schechter function is not the only
fitting function  that adequately describes the observed GSMF.  
For example a good description of the data can also be achieved by parametrizing the low mass component of the GSMF with a power-law with a more
than exponential cut-off. In general no conclusion can be drawn confidently on the GSMF shape in the region where low mass an high
mass component overlap (see also discussion in \citealt{Drory:2009p22}).\\
Indeed other descriptions of the GSMF have been proposed in the literature. 
For example \citet{Yang:2009p343} describe the GSMF of SDSS-DR7 galaxy groups as a double power law, and find it fitting adequately the GSMF of optical groups found in SDSS-DR7 data. This model is defined as:
\begin{equation}\label{eq:MF_fit}
\Phi(M_{\rm stellar}) = \Phi_0  \frac {(M_{\rm  stellar}/M_0)^\alpha }
{(x_0+(M_{\rm stellar}/M_0)^4)^\beta }  \,.
\end{equation}
We fit our data with this model: this model can adequately describe the total GSMF of COSMOS groups but we find it to be a worse description of the passive GSMF of low mass groups than the adopted double Schechter function, being unable to describe its dip at intermediate masses.

\renewcommand{\arraystretch}{1.5}
\begin{table*}[!h]
\caption[width=\textwidth]{\large Schechter Best Fit Parameters to the GSMF of Passive Galaxies}
\label{tab_param_etg}
\begin{center}
\setlength\tablinesep{3pt}
\begin{tabular}{cccccccc}
\hline
\hline
   z & V [Mpc$^{3}$] &$\phi_{1}$ &  log(\ms$_{1}$ h$^{-2}$)&$\alpha_{1}$\tablefootmark{a}&$\phi_{2}$ &  log(\ms$_{2}$ h$^{-2}$)\tablefootmark{b}&$\alpha_{2}$\tablefootmark{c} \\
\hline
\multicolumn{8}{c}{LOW MASS GROUPS}\\
\hline
0.2-0.4& 34.28&53.44$^{14.00}_{15.00}$&10.63$^{0.10}_{0.15}$&-0.35&11.50$^{6.50}_{5.00}$& 10.82&-1.26$^{0.07}_{0.10}$\\
0.4-0.6& 16.81&  46.14$^{8.00}_{8.00}$&10.67$^{0.08}_{0.06}$&-0.40& 4.68$^{0.90}_{0.85}$& 10.92&                -1.26\\
0.6-0.8& 25.20&82.94$^{10.00}_{10.00}$&10.68$^{0.06}_{0.06}$&-0.40& 1.48$^{1.20}_{1.25}$& 10.89&                -1.26\\
0.8-1.0& 15.63& 75.66$^{2.00}_{22.00}$&10.60$^{0.06}_{0.02}$&-0.40&- & - &                - \\

\hline
\multicolumn{8}{c}{HIGH MASS GROUPS}\\
\hline
0.2-0.4& 11.30&32.85$^{8.00}_{8.00}$&10.67$^{0.12}_{0.10}$&-0.99&--&--&--\\
0.8-1.0& 16.80&67.88$^{18.00}_{18.00}$&10.64$^{0.14}_{0.12}$&-0.56&--&--&--\\

\hline
\multicolumn{8}{c}{FIELD}\\
\hline
0.2-0.4&0.56 $\times$ 10$^{6}$&447.17$^{19.00}_{18.00}$&10.44$^{0.02}_{0.02}$&-0.38&2.54$^{1.02}_{0.78}$&10.67&-1.88$^{0.06}_{0.06}$\\
0.4-0.6&1.25 $\times$ 10$^{6}$&521.47$^{18.00}_{16.00}$&10.54$^{0.02}_{0.02}$&-0.40&1.53$^{0.15}_{0.15}$&10.71&                -1.88\\
0.6-0.8&1.97 $\times$ 10$^{6}$&799.49$^{20.00}_{20.00}$&10.59$^{0.02}_{0.02}$&-0.40&                  --&--&                   --\\
0.8-1.0&2.62 $\times$ 10$^{6}$&1170.54$^{28.00}_{28.00}$&10.53$^{0.02}_{0.02}$&-0.40&                  --&--&                   --\\

\hline
\hline
\end{tabular} 
\tablefoot{
\tablefoottext{a}{Fixed to -0.4 at z$>$0.4.\\}
\tablefoottext{b}{Fixed to the value of \ms$_{\alpha=-1.4}$ found for star forming galaxies.\\}
\tablefoottext{c}{Fixed to the low redshift value at z$>$0.4.\\}
}
\end{center}
\end{table*}

\section{Baryon Fraction in Star Forming and Passive Galaxies}\label{baryon_fraction}
\begin{figure}
\centering
\includegraphics[width=\columnwidth]{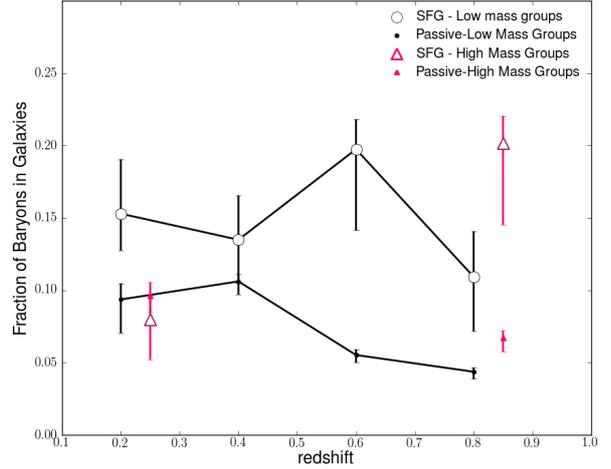}
\caption{Fraction of baryons in galaxies estimated as the integral of the GSMF over the total mass times the cosmic baryon fraction. Error bars are estimated from Montecarlo iterations over the confidence limits of the best fitting parameters in the GSMF. Empty symbols represent star forming galaxies and, the filled ones passive galaxies. Circles represent low mass groups, triangles high mass groups.}
\label{star_frac}
\end{figure}

We integrate the GSMFs down to 10$^{9}$ \Msun to obtain an estimate of the total stellar mass in passive and star forming galaxies in different environments. We compare this 
quantity with the total amount of baryons available, estimated as M$_{200}\times$f$_{b}$, where f$_{b}$ is the baryonic fraction from WMAP7 (Dunkley et al. 2008). 
The resulting quantity represents the fraction of baryons in galaxies, which is an indication on how efficiently the conversion of baryons into stars acts  as a function of redshift and total halo mass (similarly as in \citealt{Mandelbaum:2006p2567}). Figure \ref{star_frac} reveals many interesting trends in the relation between baryon conversion efficiency and galaxy properties.\\
We  find that star forming galaxies in low mass groups have the highest conversion efficiency: when compared to passive galaxies in the same environments the difference amounts to at least a factor of two at all redshifts. Interestingly, independent measures of the baryonic conversion efficiency through galaxy-galaxy lensing find a similar result (\citealt{Mandelbaum:2006p2567}). \\ We also note that star forming galaxies in high mass groups show a much lower contribution to the baryon fraction than the low mass ones, at low redshift. Indeed the fraction of baryons in star forming galaxies in high mass groups is roughly a factor of 1.5 lower than in low mass groups: although the significance of this result is less than a two sigma, it is suggested that in more massive halos the fraction of baryons locked up in star forming galaxies is lower. A similar result is found in RCS2 using galaxy-galaxy lensing \citep{vanUitert:2011p2963}.\\
Conversely the amount of baryons in passive galaxies, appears to be similar in low mass and high mass groups, being already set at redshift 0.8--1.0.
The difference between the fraction of baryons in star forming and passive galaxies holds (if not increased) also at higher redshift, being set already at redshift 0.8. On the other hand the difference in the contribution of star forming galaxies to the baryonic fraction seems to be smaller at high redshifts, where the two values are more similar due to an increased amount of baryons locked in star forming galaxies in high redshift massive groups. \\

\section{The average stellar mass fraction in groups: comparison with previous work}\label{check_sf}
\begin{figure}
\centering
\includegraphics[width=\columnwidth]{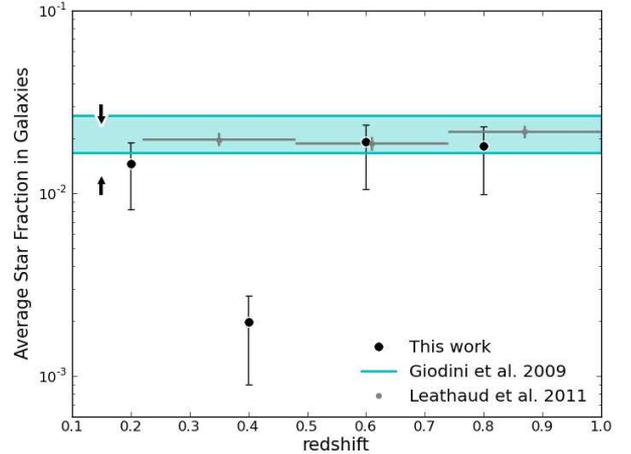}
\caption{Black points show the average stellar mass fraction for low mass COSMOS X-ray groups sample computed by summing over the composite GSMF at each redshift. These values are compared with the average stellar mass fraction presented in \citealt{Giodini:2009p923} (dashed region). Grey points are the stellar mass fraction computed by \citet{2011arXiv1109.0010L}, also on COSMOS X-ray detected groups but by using HOD analysis. The arrows show the upper and lower bounds of the systematic errors  on the stellar mass estimates at low redshift as estimated by \citet{2011arXiv1109.0010L}. }
\label{sf_test}
\end{figure}

A straightforward outcome from our analysis is the average stellar mass fraction in groups in the redshift/total mass bin described in the previous sections. We can quickly compute the average stellar mass fraction within R$_{200}$ by summing over the groups' total GSMF down to the completeness mass for each redshift bin and correcting these value for the statistical contribution of lower mass galaxies (1$\%$ at z$<$0.5 and 9$\%$ at 0.5$<z<$1.0; \citealt{Giodini:2009p923}). We then divide the ensuing number the summed M$_{200}$ of the groups considered. These values are showed as black points in Figure \ref{sf_test}\footnote{Note that these values cannot be directly compared with those in Figure \ref{star_frac} since those are computed by integrating the mass function down to a different stellar mass.}. 
As a consistency check we compare our average stellar mass fraction with those published in \citet{Giodini:2009p923} for low mass groups. We correct these values for a Chabrier IMF and for the additional systematic shift by $\sim$0.2-0.4 dex between K-band and SED computed stellar masses discussed in \citet{Ilbert:2010p420} (we use 0.3 dex as an average value). \\
We also compare our results with the recently published average groups' stellar mass fractions by \citet{2011arXiv1109.0010L} : these authors constrain the fraction of stars in group-sized haloes by using a statistical Halo Occupation Distribution model that jointly constrained by data from lensing, clustering, and the stellar mass function.
In Figure \ref{sf_test}, we show the comparison of the values of average stellar mass fraction computed from the composite GSMF for low mass groups (M$_{200}<$6$\times$10$^{13}$M$_{\odot}$) to that found in \citet{Giodini:2009p923} and \citet{2011arXiv1109.0010L} for similar average total masses (we approximate the difference between M$_{200}$ and M$_{500}$ to 30$\%$, which corresponds to the difference for NFW haloes with concentration equal 5).  
\footnote{Since the difference in the median mass between low and high mass groups sample is only a factor of 2-4 (at low-high redshift, respectively), the comparison looks very similar to that at low redshift and we decide not to show these points on the plot for clarity.}\\
Reassuringly, Figure \ref{sf_test} shows a broad agreement between the three measurement, even if computed with different methods, confirming the robustness of the estimated stellar mas fraction values. The significantly lower value of the average stellar mass fraction at z$\sim$0.4 is not surprising since there is a lack of massive galaxies in COSMOS at this redshift. (see also \citealt{Pannella:2009p3946}).
\section{The sSFR in galaxy groups}\label{ssfr}
\begin{figure*}
\centering
\includegraphics[width=\textwidth]{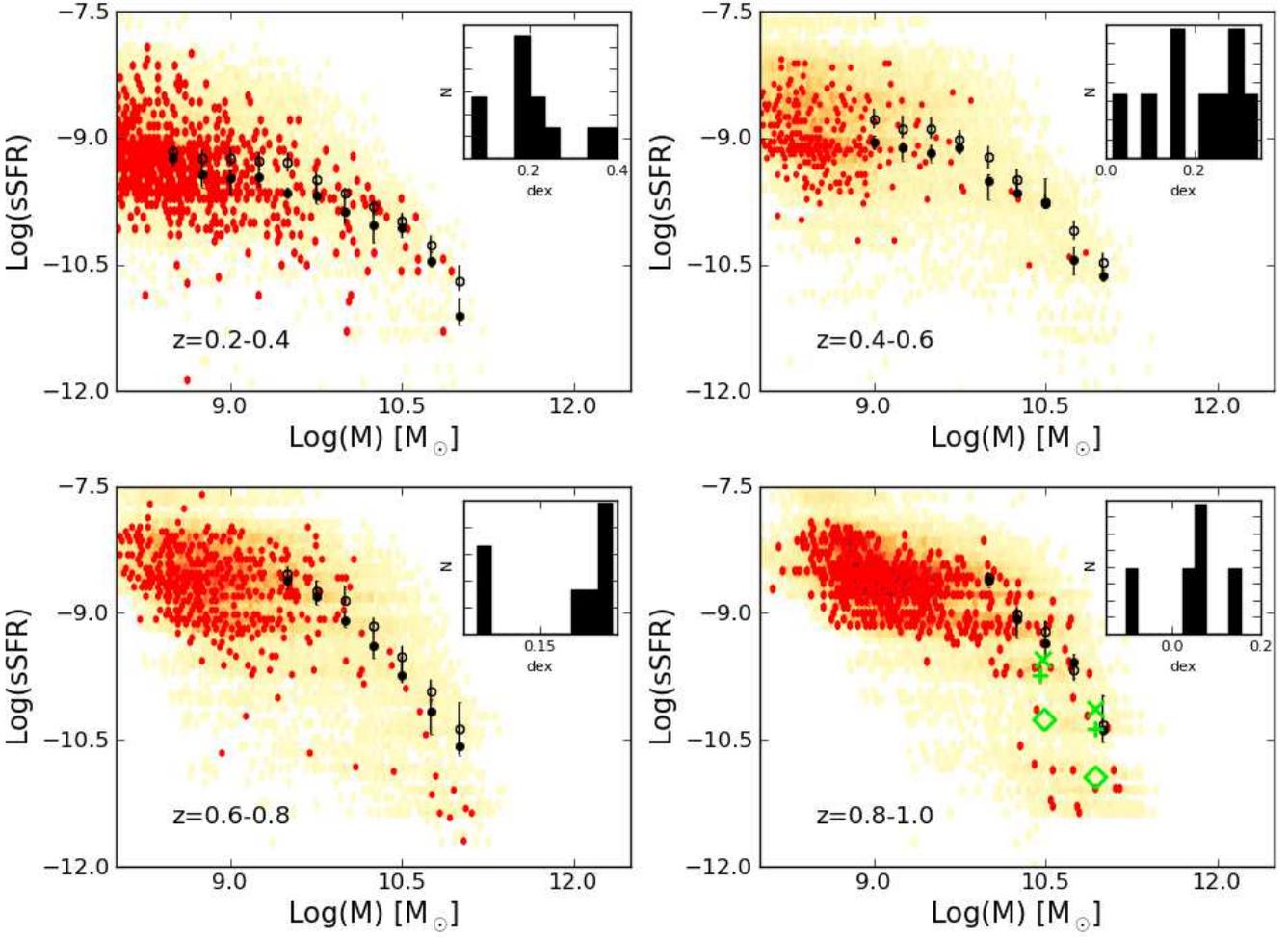}
\caption{Specific star formation rate versus stellar mass for star forming galaxies in groups (red) and the field (yellow). Individual points are re-binned in hexagons and only bins with more than two counts are shown.  Black filled points show the median for the groups, empty ones for the field. Inset histograms show the distribution of the difference between the median relation in groups and in the field.
Green symbols at z=0.8-1.0 are obtained from MIPS stacking by Patel et al. (2009) and represent the field (crosses), groups (plus sign) and clusters (diamonds).}
\label{ssfr_mass}
  \end{figure*}

\begin{figure}
\centering
\includegraphics[width=\columnwidth]{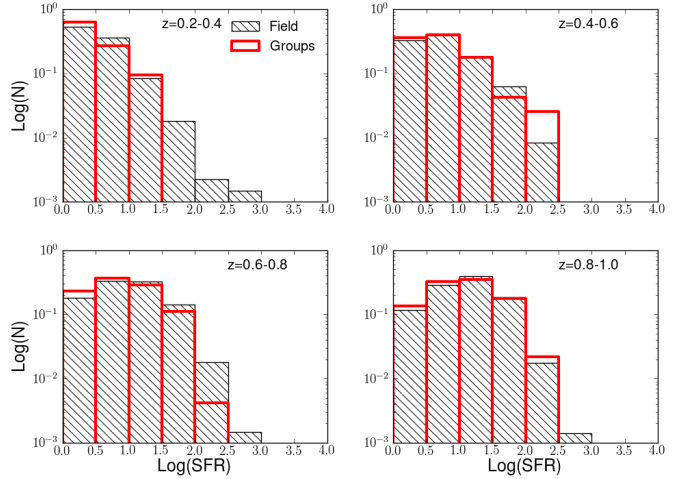}
\caption{Distribution of SFRs for highly star forming galaxies (SFR$>$ 1\Msun y$^{-1}$) in groups (red) and field (grey hatched bars).  }
\label{sfr_act}
 \end{figure}

In order to better understand the differences in the galaxy stellar mass function between groups and the field, we analyse the distribution of the specific star formation rate (sSFR=SFR/M, where M is the stellar mass of a galaxy).
The specific star formation rate is a side product of the SED fitting of the photometric points for each galaxy performed by Ilbert et al. (2009). 
In Figure \ref{ssfr_mass} we show the  sSFR as a function of the stellar mass for galaxies associated to groups (in red) and to the field (in yellow), while circles mark the median values of the distributions in stellar mass bins.\\ 
The distributions in the two environments cover approximately the same region in the plane; when plotting the median values in bins of mass the median sSFR in the field (empty circles) is higher than in groups (solid circles).  However, even if the difference between the two median values is not significant at all stellar masses, when comparing the individual values, it is significant that the values of median SSFR in groups are consistently lower than those in the field. The inset histograms in Figure \ref{ssfr_mass} show the distribution of the difference in dex between these median values: at low redshift the typical discrepancy amounts to $\sim$0.2 dex.\\
This difference is not unexpected since it has already been shown that the median sSFR, at least in massive galaxies, declines as a function of local density (Kauffman at al 2004; Patel et al. 2009). The difference we find is similar to that found using SFR measurement from MIPS sources in different environment (Patel et al. 2009; green points in Fig. \ref{ssfr_mass}).\\
Our finding suggests that the population of star forming galaxies is somehow modified in its capability of forming stars, by the presence of a surrounding environment. To strengthen this point we perform a KS test on the data--points to test if the sSFRs in groups and field are consistent with being drawn from the same distribution. We only consider points above the highest completeness mass. The associated probabilities  are below 1$\%$ at all redshifts, suggesting that the two distributions are different at a high level of significance. 
Indeed if we compute the normalized cumulative sSFR distribution in the different environments and compare the median sSFR above the completeness mass,
this value is lower for groups than field. This means that low mass groups show an excess of low-sSFR galaxies.\\ 
These results suggests that the distribution of sSFR is strongly modified when a galaxy enters in a structure, in agreement with recent findings of a `reduced star-formation' galaxy population in groups and clusters\citep{Vulcani:2010p1322, Balogh:2010p1150}.  \\
Having a large range of redshift for both field and groups environment, we can point out that the field distribution at redshifts 0.2--0.4 resembles that of groups at z=0.6--0.8 by matching the median of their cumulative sSFR distributions. If the difference in sSFR distribution corresponds an age difference, it suggests a delay by $\sim$3 Gyr between field and groups. A similar delay between higher and lower density environments has been quoted by \citet{Bolzonella:2009p425}.  \\
Finally Figure \ref{sfr_act} shows the distribution of highly star forming ($>$1\Msun y$^{-1}$) galaxies in groups (red histogram) and the field (line-filled histogram). 
The two distributions do not exhibit significant differences at z$>$0.4, while at lower redshift  we note that the galaxies with higher star formation rates ($>$30 \Msun y$^{-1}$) are completely absent in X-ray selected groups.


\section{Discussion}\label{discussion}
The analysis of the galaxy stellar mass function revealed intriguing differences between groups and the field. Interpreting these differences can tell us how the build up of the stellar mass in galaxies (which is a local process) is affected by global properties of the parent halo. In the following sections we discuss our results.

\subsection{Star Forming Galaxies}
In our analysis we show that the stellar mass distribution of star forming galaxies can be described as a single Schechter function at all redshifts and in all the environments examined in this work (field, high and low mass groups).  However, at z=0.2--0.4 low-mass groups exhibit a galaxy stellar mass distribution where the characteristic mass tends to be larger and the slope less negative than the corresponding parameters of the field GSMF.
The tendency towards a 50$\%$ larger characteristic mass of the star forming galaxies in low-mass groups with respect to the field holds when the slope of the star forming GSMF is set to be equal to -1.4. The same is not seen in larger systems, where the fraction of red galaxies increases at a given mass: it is likely that such galaxies have been quenched more efficiently in massive groups, explaining the difference in characteristic stellar mass for star forming galaxies between these two environments. \\ 
A possible explanation for the presence of massive star forming galaxies in groups may be that the quenching is proceeding at a slower pace in groups  than in massive clusters, enabling these systems to retain for a longer time the necessary fuel for star formation.
Interaction with the surrounding hot gas in groups may lead to rejuvenation of the star formation (Gavazzi$\&$Jaffe 1985). However, our analysis of the specific star formation rate distribution indicates that the median values are lower in these environments at least since redshift 0.8, suggesting that rejuvenation may be prevented in galaxy groups. This indication is supported by the presence of a large fraction of galaxies with intermediate color (``gren valley" galaxies) and reduced star formation rates, in galaxy groups at z$\sim$0.8 \citep{Balogh:2010p1150}. \\
Therefore, if the same process is responsible for quenching star formation in groups and more massive structures, our findings suggest that this process is downsized or slowed down in galaxy groups. As  a consequence, in the two-process scenario drawn by Peng et al. 2010, the "mass quenching", driven by feedback processes, may depend also on the environment. \\
Furthermore we find that the fraction of star forming galaxies is larger in the field than in groups, at any stellar mass. The ensuing suggestion that the field hosts a star formation activity which is more extended in time is supported by the larger values of sSFR obtained for field galaxies with respect to those in groups. 
This finding complements results obtained with smaller samples by \citet{Patel:2009p1002} and in more massive structures by \citet{Vulcani:2010p919}. Interestingly the same is not found in similar analysis performed on optically detected groups \citep{McGee:2011p2883},  which may indicate that the quenching strength is larger in the evolved and virialized X-ray detected systems. If galaxy groups significantly contribute to globally reduce the star formation, environmental effects already effective on groups' scale, such as strangulation, are important in shaping the distribution of galaxy properties in more massive structures. Furthermore our results agree with a scenario in which environment regulates the time-scales of star formation history as suggested by observations of cluster and field galaxies at high redshift \citep{Rettura:2011p3936}.\\
Another interesting feature of the characteristic mass of star forming galaxies in groups is its remarkable stability between redshift 0.2 and 1.0. 
As already suggested by Peng et al. 2010, this fact can be understood if the mass quenching proceeds at the same pace as the star formation rate. Indeed if feedback processes are responsible for the quenching of star formation in massive galaxies, this dependence is expected \citep{2007MNRAS.382..960K}. As a consequence, if the star formation rate is decreased in groups, so is the mass quenching, supporting the interpretation for the presence of more massive star forming galaxies than in the field.\\
Furthermore, our findings confirm that galaxy evolution is faster in higher density environments, as indicated by the more rapid decline in the fraction of star forming galaxies in groups than in the field. By matching the cumulative distributions of sSFR in different environments, we estimated the delay between low mass groups and field being roughly 3 Gyr. This corresponds to a growth by 3$\times$10$^{9}$ \Msun in stellar mass assuming the star formation rate of a typical galaxy as the Milky Way ($\sim$1 \Msun y$^{-1}$). This amount is much lower than the difference in characteristic mass between groups and field, indicating that the field will never reach the mass distribution of groups, where the build up of mass has been more efficient.  Consistently, only field galaxies exhibit values of the SFR that exceeds 30 M$_{\odot}$y$^{-1}$ at z$<$0.4. These results point towards a strong evolution of the SFR per unit of halo mass at z$<0.4$, in agreement with recent results from Herschel data \citep{Popesso:10p3436}. \\
A synthetic view of the previously described behaviours is offered by the redshift dependence of the star forming galaxy fraction shown in Figure \ref{frac_act} for the three environments considered in this work: there are less and less star forming galaxies as the universe ages, but the denser the environment  the lower this fraction. In other words, groups are intermediate environments when considering the Butcher-Oemler effect \citep{1995MNRAS.274..153K, 2005MNRAS.358...71W, 2007MNRAS.376.1425G}.\\
A word of caution should be spent when considering the evolution with redshift of the low mass systems examined in this paper. In fact, due to the bias towards brightest systems at higher redshifts introduced by the X-ray selection, the median mass of low mass groups at z=0.8--1.0 is a factor of two higher than that at z=0.2--0.4. Therefore in this work we do not aim at constraining the properties of the progenitors of low redshift groups, whereas at describing how groups with similar M$_{200}$ appear at different epochs. In general a higher median mass does not affect largely our results on SFR and baryonic fraction evolution, but turns our results in lower limits on the evolution. Indeed both these quantities decrease with increasing mass, diminishing the strength of the evolutionary trend. In order to confidently draw conclusions on the evolution of the GSMF a further work on a complementary sample of high redshift low mass groups from \textit{Chandra Deep Field South} is ongoing (Giodini et al. in preparation).\\
Finally our analysis highlighted environmentally dependent differences in the amount of baryons locked in stars in different types of galaxies.\\
Low mass groups show a higher fraction of star forming galaxies when compared to high mass groups and clusters. Translating this to the amount of baryons corresponding to star forming galaxies, 15$\%$ of the baryons are distributed in stars within star forming galaxies in low mass groups at z=0.2--0.4, but only 10$\%$ in higher mass systems. Thus, in spite of the fractional increase of the number of passive systems at later epochs, about two thirds of the total stellar mass of a group (excluding the intra-cluster light) is locked in star forming galaxies at z=0.3.\\
Furthermore, the large amount of baryons in star forming  galaxies  in low mass groups is already settled at redshift 0.8 and it may be evolving towards a lower value at z=0.8--1.0.\\
In \citet{Giodini:2009p923} we found that the fraction of stars per unit halo mass ($f_\mathrm{star}$) is a function of M$_{200}$ for  systems with 10$^{13}<$M$_{200}<$10$^{15}$ \Msun at z$<$1. Interestingly, the difference in $f_\mathrm{star}$ between groups of M$_{200}$ corresponding to the median for low and high mass groups used in this work is comparable with that in the fraction of baryons locked in star forming galaxies. This suggests that a relative excess of baryons in star forming galaxies within low mass groups can explain the difference in $f_\mathrm{star}$in groups of different M$_{200}$.\\
At a first glance we could try to explain the higher fraction of baryons with a recent infall of star forming galaxies in low mass groups. If so we would expect that the total stellar mass in star forming galaxies does not dependent on the total mass of the structure (i.e. on its volume), but on the collecting area. Assuming that the total stellar mass of passive galaxies is a good tracer of the total mass of the system, we  expect that $\frac{M^\mathrm{SFG}}{M^\mathrm{passive}}\propto\frac{1}{R_\mathrm{200}}$. 
However we find that the average growth  of the amount of baryons locked in star forming galaxies as a function of M$_{200}$ is faster than that predicted using the median mass of our group subsamples (a factor of 2 against 1.5). This leads us to the conclusion that galaxy groups of $\sim$3$\times$10$^{13}$ \Msun may have a more efficient conversion of baryons into stars.  A first tantalizing evidence of an enhanced baryonic conversion efficiency in galaxy groups was found by \citet{Mcgaugh:1p2899}. We confirm this result and in addition show that the excess baryons are locked in star-forming galaxies. \\
It is worth noticing that the enhanced baryonic conversion efficiency in less massive galaxy systems likely has a higher metal enrichment as a consequence. Interestingly,
this is found at the mass regime of clusters in the local Universe \citep{2011arXiv1109.0390Z}.

\subsection{Passive Galaxies}
The analysis of the data presented in this paper show the existence of passively evolving galaxies with stellar masses larger than 10$^{10}$\Msun up to redshift 1, whatever the surrounding environment (i.e. both in the field, high and low mass groups). This is a signature of an origin likely not connected with the large scale environment, but with processes internal to massive galaxies (the so called ``feedback").\\
These passive galaxies dominate the whole population of galaxies with stellar mass larger than 10$^{10}$\Msun at all redshifts in groups, while in the field they become dominant only at z$<$0.6.\\
On the other hand, the population of low mass passive galaxies (M$<$10$^{10}$\Msun)  builds up since z=1 in a continuous fashion, both in groups and in the field. Interestingly, the difference  in the GSMF of passive galaxies between groups and field decreases by a factor of 2--3 from z=0.8--1.0 to z=0.2--0.4, indicating that field and groups were increasingly similar at higher redshift. At the same time, the number density of passively evolving galaxies with stellar masses in the range of 2$\times$10$^{9}$--2$\times$10$^{10}$ \Msun is lower in the field than in groups. This difference increases when more massive groups are considered, which suggests the existence of a process of ``secular quenching" that depends both on environment and stellar mass. \\
Therefore, the GSMF of passive galaxies cannot be described by a single Schechter function whenever the data extend to Log(M/M$_{\odot}$)$<$10 (at z$<$0.8). In this case a suitable fit is achieved by using a double Schechter function, where the second component is sensible to the behaviour of lower mass galaxies.\\
When using this parametrization, we find the slope of the low mass component of the mass function for passive galaxies to be different between the groups and the field. This slope is quite steep ($\sim$-1.8) in the field at odds with predictions based on SDSS and zCOSMOS data \citep{Peng:2010p568}, according to which it should be as steep as that of star forming galaxies. The groups GSMF, instead, behaves as prescribed by the predictions, with a low mass slope that is compatible with that of star forming galaxies.\\
Interestingly, a hint for a rising slope at low mass can be seen in \citet{Ilbert:2010p420} (their Fig. 11), despite the higher cut in stellar mass due to the selection according to IRAC photometry. The latter authors divide their sample according to morphology: interestingly the rising slope disappears for quiescent galaxies with an elliptical morphology. Therefore the low mass upturn is likely to be associated with quenched galaxies which are not ellipticals. These galaxies are connected to environmental processes already active in groups such as gas stripping and starvation caused by a diffuse intra-group medium, or harassment.
Indeed it is likely that the low mass component of the GSMF stems from quenching of satellite galaxies due to environmental effect (``environmental quenching"), while  the high mass component is subject to events that shut off star formation via feedback processes (``mass--quenching").  The model of galaxy evolution suggested by \citep{Peng:2010p568} predict the low mass slope to be the same in all environments, since environmental quenching is assumed to be independent on the halo mass.\\
The presence of an upturn in our data is likely enhanced by the lack of intermediate mass galaxies in the field. This effect produces a ``wiggle" in the GSMF between 10$^{9}$-10$^{10}$ \Msun , which may be explained by a delayed appearance of  passive galaxies of this mass in the field.  
 A similar interpretation has been proposed by Tanaka et al. (2005) in a study on the build up of the faint end of the red sequence as a function of the environment. They found that the ratio of giant (Log(M/\Msun)$>$10.6) to dwarf (9.7$<$Log(M/\Msun)$<$10.6) red sequence galaxies is larger in the field at any redshift, indicating that such faint galaxies are relative rare in the field.\\
Therefore our finding suggest that the build up of the the secondary quenched component is independent of the stellar mass in groups, as witnessed by the flatness of the slope of the GSMF, whereas in the field the quenching of low mass galaxies depends on their stellar mass. In particular the very steep slope found at Log(M/\Msun)$<$9.5 indicates a preferential  quenching of such galaxies.
Interestingly, semi--analytic models implementing the latest recipes of galaxy evolution fails to reproduce the fraction of dwarf passive satellites in structures, leading to an overproduction of  low mass quenched galaxies \citep[e.g.]{2011arXiv1105.0674W}. Current models struggle in justifying the destruction of these low mass galaxies via environmental processes, however the introduction of an environmental dependence in the quenching process, as suggested by our data, may alleviate this problem. \\

\section{Conclusions}\label{conclusions}
In this paper we investigate the distribution of stellar mass in galaxies within X-ray detected galaxy groups in the COSMOS survey. \\
After building the composite distributions for a sample of 160 groups divided into two sub-samples of high and low mass groups, we investigate the shape of the distribution for passive and star forming galaxies, comparing it to that of the field.  \\
Our analysis sheds light on how  the transition between star-forming and passive galaxies occurs in different environments. In particular, we highlight how the field builds up at low redshift a population of low mass (M$<$10$^{9.5}$ \Msun) quenched galaxies which does not appear in groups and we unveil the slower build up of an intermediate mass (10$^{9.5}<$M$<$10$^{10.5}$ \Msun) quenched component in the field. As a consequence the distribution of stellar mass for passive galaxies shows differences in the shape between the groups and the field.\\
On the other hand, the stellar mass distribution of star forming galaxies is similar in the shape in all the environments and can be adequately described by a single Schechter function. However, we find indication for the bulk of the stellar mass in star forming galaxies being more massive in low mass groups then in high mass groups at low redshift (M$_\mathrm{200}<$6$\times$10$^{13}$), and we interpret this as the the quenching process acting at a slower pace. \\
More generally we find that the distribution of sSFR is different between X-ray detected groups and the field, with groups showing median star formation rates lower than the field at all stellar masses, and we estimate the delay between field and structures to be $\sim$ 3 Gyr. Accordingly the fraction of star forming galaxies in groups is lower than in the field at all redshifts, with low mass groups being intermediate between field and more massive systems. 
In general the significance of the above findings decreases at high redshifts, suggesting that groups and field may be sharing more similar properties at z$\sim$1.\\
Finally, we find that despite the increase of the passive population at lower redshift, at z=0.2-0.4 groups have two thirds of their stellar mass locked in star forming galaxies and low mass groups exhibit a larger fraction of baryons in star forming galaxies, in agreement with recent findings suggesting that that these systems convert more efficiently baryons into stars.
\section*{Acknowledgements}
SG acknowledges support from the Netherlands Organization for Scientific Research (NWO) through a VIDI grant. We acknowledge the contributions of the entire COSMOS collaboration; more informations on the COSMOS survey are available at {http://www.astr.caltech.edu/$\sim$cosmos}. SG acknowledge H. Boehringer for having contributed to the start of this project and H. Hoekstra, G. Guzzo, J. Brinchmann, S. Weinmann, D. Capozzi, T. Ponman, B. Vulcani and A. Leauthaud for helpful discussion. DP acknowledges the kind hospitality of the MPE.
\bibliography{giodini_v2_paper}

\begin{thebibliography}{67}
\expandafter\ifx\csname natexlab\endcsname\relax\def\natexlab#1{#1}\fi

\bibitem[{Baldry {et~al.}(2004)Baldry, Glazebrook, Brinkmann, Ivezi{\'c},
  Lupton, Nichol, \& Szalay}]{Baldry:2004p4667}
Baldry, I.~K., Glazebrook, K., Brinkmann, J., {et~al.} 2004, The Astrophysical
  Journal, 600, 681

\bibitem[{Baldry {et~al.}(2008)Baldry, Glazebrook, \& Driver}]{Baldry:2008p58}
Baldry, I.~K., Glazebrook, K., \& Driver, S.~P. 2008, Monthly Notices of the
  Royal Astronomical Society, 388, 945

\bibitem[{Balogh {et~al.}(2001)Balogh, Christlein, Zabludoff, \&
  Zaritsky}]{Balogh:2001p86}
Balogh, M.~L., Christlein, D., Zabludoff, A.~I., \& Zaritsky, D. 2001, The
  Astrophysical Journal, 557, 117

\bibitem[{Balogh {et~al.}(2010)Balogh, McGee, Wilman, Finoguenov, Parker,
  Connelly, Mulchaey, Bower, Tanaka, \& Giodini}]{Balogh:2010p1150}
Balogh, M.~L., McGee, S.~L., Wilman, D.~J., {et~al.} 2010, eprint arXiv, 1011,
  5509

\bibitem[{Blanton {et~al.}(2005)Blanton, Lupton, Schlegel, Strauss, Brinkmann,
  Fukugita, \& Loveday}]{Blanton:2005p4751}
Blanton, M.~R., Lupton, R.~H., Schlegel, D.~J., {et~al.} 2005, The
  Astrophysical Journal, 631, 208

\bibitem[{Bolzonella {et~al.}(2010)Bolzonella, Kovac, Pozzetti, Zucca,
  Cucciati, Lilly, Peng, Iovino, Zamorani, Vergani, Tasca, Lamareille, Oesch,
  Caputi, Kampczyk, Bardelli, Maier, Abbas, Knobel, Scodeggio, Carollo,
  Contini, Kneib, Fevre, Mainieri, Renzini, Bongiorno, Coppa, de~la Torre,
  de~Ravel, Franzetti, Garilli, Borgne, Brun, Mignoli, Pello, Perez-Montero,
  Ricciardelli, Silverman, Tanaka, Tresse, Bottini, Cappi, Cassata, Cimatti,
  Guzzo, Koekemoer, Leauthaud, Maccagni, Marinoni, McCracken, Memeo, Meneux,
  Porciani, Scaramella, Aussel, Capak, Halliday, Ilbert, Kartaltepe, Salvato,
  Sanders, Scarlata, Scoville, Taniguchi, \& Thompson}]{Bolzonella:2009p425}
Bolzonella, M., Kovac, K., Pozzetti, L., {et~al.} 2010, eprint arXiv, 0907, 13

\bibitem[{{Bundy} {et~al.}(2006){Bundy}, {Ellis}, {Conselice}, {Taylor},
  {Cooper}, {Willmer}, {Weiner}, {Coil}, {Noeske}, \&
  {Eisenhardt}}]{2006ApJ...651..120B}
{Bundy}, K., {Ellis}, R.~S., {Conselice}, C.~J., {et~al.} 2006, \apj, 651, 120

\bibitem[{{Butcher} \& {Oemler}(1978)}]{1978ApJ...219...18B}
{Butcher}, H. \& {Oemler}, Jr., A. 1978, \apj, 219, 18

\bibitem[{Capak {et~al.}(2007)Capak, Aussel, Ajiki, McCracken, Mobasher,
  Scoville, Shopbell, Taniguchi, Thompson, Tribiano, Sasaki, Blain, Brusa,
  Carilli, Comastri, Carollo, Cassata, Colbert, Ellis, Elvis, Giavalisco,
  Green, Guzzo, Hasinger, Ilbert, Impey, Jahnke, Kartaltepe, Kneib, Koda,
  Koekemoer, Komiyama, Leauthaud, Fevre, Lilly, Liu, Massey, Miyazaki,
  Murayama, Nagao, Peacock, Pickles, Porciani, Renzini, Rhodes, Rich, Salvato,
  Sanders, Scarlata, Schiminovich, Schinnerer, Scodeggio, Sheth, Shioya, Tasca,
  Taylor, Yan, \& Zamorani}]{Capak:2007p850}
Capak, P., Aussel, H., Ajiki, M., {et~al.} 2007, The Astrophysical Journal
  Supplement Series, 172, 99

\bibitem[{Chabrier(2003)}]{Chabrier:2003p4062}
Chabrier, G. 2003, The Publications of the Astronomical Society of the Pacific,
  115, 763

\bibitem[{Christlein(2000)}]{Christlein:2000p4169}
Christlein, D. 2000, The Astrophysical Journal, 545, 145

\bibitem[{Crain {et~al.}(2009)Crain, Theuns, Vecchia, Eke, Frenk, Jenkins, Kay,
  Peacock, Pearce, Schaye, Springel, Thomas, White, \&
  Wiersma}]{Crain:2009p1881}
Crain, R.~A., Theuns, T., Vecchia, C.~D., {et~al.} 2009, Monthly Notices of the
  Royal Astronomical Society, 399, 1773

\bibitem[{{Croton} {et~al.}(2006){Croton}, {Springel}, {White}, {De Lucia},
  {Frenk}, {Gao}, {Jenkins}, {Kauffmann}, {Navarro}, \&
  {Yoshida}}]{2006MNRAS.365...11C}
{Croton}, D.~J., {Springel}, V., {White}, S.~D.~M., {et~al.} 2006, \mnras, 365,
  11

\bibitem[{Donahue {et~al.}(2010)Donahue, Bruch, Wang, Voit, Hicks, Haarsma,
  Croston, Pratt, Pierini, O'Connell, \& B{\"o}hringer}]{Donahue:2010p2318}
Donahue, M., Bruch, S., Wang, E., {et~al.} 2010, The Astrophysical Journal,
  715, 881

\bibitem[{Driver {et~al.}(1994)Driver, Phillipps, Davies, Morgan, \&
  Disney}]{Driver:1994p1615}
Driver, S., Phillipps, S., Davies, J., Morgan, I., \& Disney, M. 1994, Dwarf
  Galaxies, 49, 141

\bibitem[{Drory {et~al.}(2009)Drory, Bundy, Leauthaud, Scoville, Capak, Ilbert,
  Kartaltepe, Kneib, McCracken, Salvato, Sanders, Thompson, \&
  Willott}]{Drory:2009p22}
Drory, N., Bundy, K., Leauthaud, A., {et~al.} 2009, The Astrophysical Journal,
  707, 1595

\bibitem[{{Elbaz} {et~al.}(2007){Elbaz}, {Daddi}, {Le Borgne}, {Dickinson},
  {Alexander}, {Chary}, {Starck}, {Brandt}, {Kitzbichler}, {MacDonald},
  {Nonino}, {Popesso}, {Stern}, \& {Vanzella}}]{2007A&A...468...33E}
{Elbaz}, D., {Daddi}, E., {Le Borgne}, D., {et~al.} 2007, \aap, 468, 33

\bibitem[{{Finoguenov} {et~al.}(2009){Finoguenov}, {Connelly}, {Parker},
  {Wilman}, {Mulchaey}, {Saglia}, {Balogh}, {Bower}, \&
  {McGee}}]{2009ApJ...704..564F}
{Finoguenov}, A., {Connelly}, J.~L., {Parker}, L.~C., {et~al.} 2009, \apj, 704,
  564

\bibitem[{Finoguenov {et~al.}(2007)Finoguenov, Guzzo, Hasinger, Scoville,
  Aussel, B{\"o}hringer, Brusa, Capak, Cappelluti, Comastri, Giodini,
  Griffiths, Impey, Koekemoer, Kneib, Leauthaud, F{\`e}vre, Lilly, Mainieri,
  Massey, McCracken, Mobasher, Murayama, Peacock, Sakelliou, Schinnerer,
  Silverman, Smol{\v c}i{\'c}, Taniguchi, Tasca, Taylor, Trump, \&
  Zamorani}]{Finoguenov:2007p475}
Finoguenov, A., Guzzo, L., Hasinger, G., {et~al.} 2007, The Astrophysical
  Journal Supplement Series, 172, 182

\bibitem[{Finoguenov {et~al.}(2010)Finoguenov, Watson, Tanaka, Simpson,
  Cirasuolo, Dunlop, Peacock, Farrah, Akiyama, Ueda, Smol{\v c}i{\'c}, Stewart,
  Rawlings, van Breukelen, Almaini, Clewley, Bonfield, Jarvis, Barr, Foucaud,
  McLure, Sekiguchi, \& Egami}]{Finoguenov:2010p2012}
Finoguenov, A., Watson, M.~G., Tanaka, M., {et~al.} 2010, Monthly Notices of
  the Royal Astronomical Society, 403, 2063

\bibitem[{Gehrels(1986)}]{Gehrels:1986p698}
Gehrels, N. 1986, Astrophysical Journal, 303, 336

\bibitem[{George {et~al.}(2011)George, Leauthaud, Bundy, Finoguenov, Tinker,
  Lin, Mei, Kneib, Aussel, Behroozi, Busha, Capak, Coccato, Covone, Faure,
  Fiorenza, Ilbert, Floc'h, Koekemoer, Tanaka, Wechsler, \&
  Wolk}]{George:2011p3935}
George, M.~R., Leauthaud, A., Bundy, K., {et~al.} 2011, eprint arXiv, 1109,
  6040

\bibitem[{{Gerke} {et~al.}(2007){Gerke}, {Newman}, {Faber}, {Cooper}, {Croton},
  {Davis}, {Willmer}, {Yan}, {Coil}, {Guhathakurta}, {Koo}, \&
  {Weiner}}]{2007MNRAS.376.1425G}
{Gerke}, B.~F., {Newman}, J.~A., {Faber}, S.~M., {et~al.} 2007, \mnras, 376,
  1425

\bibitem[{Giodini {et~al.}(2009)Giodini, Pierini, Finoguenov, Pratt,
  Boehringer, Leauthaud, Guzzo, Aussel, Bolzonella, Capak, Elvis, Hasinger,
  Ilbert, Kartaltepe, Koekemoer, Lilly, Massey, McCracken, Rhodes, Salvato,
  Sanders, Scoville, Sasaki, Smolcic, Taniguchi, Thompson, \& the
  COSMOS~Collaboration}]{Giodini:2009p923}
Giodini, S., Pierini, D., Finoguenov, A., {et~al.} 2009, The Astrophysical
  Journal, 703, 982

\bibitem[{Gonz{\'a}lez {et~al.}(2006)Gonz{\'a}lez, Lares, Lambas, \&
  Valotto}]{Gonzalez:2006p4487}
Gonz{\'a}lez, R.~E., Lares, M., Lambas, D.~G., \& Valotto, C. 2006, Astronomy
  and Astrophysics, 445, 51

\bibitem[{Goto {et~al.}(2003)Goto, Okamura, Yagi, Sheth, Bahcall, Zabel,
  Crouch, Sekiguchi, Annis, Bernardi, Chong, G{\'o}mez, Hansen, Kim, Knudson,
  McKay, \& Miller}]{Goto:2003p3583}
Goto, T., Okamura, S., Yagi, M., {et~al.} 2003, Publications of the
  Astronomical Society of Japan, 55, 739

\bibitem[{Hasinger {et~al.}(2007)Hasinger, Cappelluti, Brunner, Brusa,
  Comastri, Elvis, Finoguenov, Fiore, Franceschini, Gilli, Griffiths, Lehmann,
  Mainieri, Matt, Matute, Miyaji, Molendi, Paltani, Sanders, Scoville, Tresse,
  Urry, Vettolani, \& Zamorani}]{Hasinger:2007p1957}
Hasinger, G., Cappelluti, N., Brunner, H., {et~al.} 2007, The Astrophysical
  Journal Supplement Series, 172, 29

\bibitem[{Hilton {et~al.}(2005)Hilton, Collins, de~Propris, Baldry, Baugh,
  Bland-Hawthorn, Bridges, Cannon, Cole, Colless, Couch, Dalton, Driver,
  Efstathiou, Ellis, Frenk, Glazebrook, Jackson, Lahav, Lewis, Lumsden, Maddox,
  Madgwick, Norberg, Peacock, Peterson, Sutherland, \&
  Taylor}]{Hilton:2005p4608}
Hilton, M., Collins, C., de~Propris, R., {et~al.} 2005, Monthly Notices of the
  Royal Astronomical Society, 363, 661

\bibitem[{Ilbert {et~al.}(2009)Ilbert, Capak, Salvato, Aussel, McCracken,
  Sanders, Scoville, Kartaltepe, Arnouts, Floc'h, Mobasher, Taniguchi,
  Lamareille, Leauthaud, Sasaki, Thompson, Zamojski, Zamorani, Bardelli,
  Bolzonella, Bongiorno, Brusa, Caputi, Carollo, Contini, Cook, Coppa,
  Cucciati, de~la Torre, de~Ravel, Franzetti, Garilli, Hasinger, Iovino,
  Kampczyk, Kneib, Knobel, Kovac, Borgne, Brun, F{\`e}vre, Lilly, Looper,
  Maier, Mainieri, Mellier, Mignoli, Murayama, Pell{\`o}, Peng,
  P{\'e}rez-Montero, Renzini, Ricciardelli, Schiminovich, Scodeggio, Shioya,
  Silverman, Surace, Tanaka, Tasca, Tresse, Vergani, \&
  Zucca}]{Ilbert:2009p917}
Ilbert, O., Capak, P., Salvato, M., {et~al.} 2009, The Astrophysical Journal,
  690, 1236

\bibitem[{Ilbert {et~al.}(2010)Ilbert, Salvato, Floc'h, Aussel, Capak,
  McCracken, Mobasher, Kartaltepe, Scoville, Sanders, Arnouts, Bundy, Cassata,
  Kneib, Koekemoer, F{\`e}vre, Lilly, Surace, Taniguchi, Tasca, Thompson,
  Tresse, Zamojski, Zamorani, \& Zucca}]{Ilbert:2010p420}
Ilbert, O., Salvato, M., Floc'h, E.~L., {et~al.} 2010, The Astrophysical
  Journal, 709, 644

\bibitem[{Juneau {et~al.}(2005)Juneau, Glazebrook, Crampton, McCarthy,
  Savaglio, Abraham, Carlberg, Chen, Borgne, Marzke, Roth, J{\o}rgensen, Hook,
  \& Murowinski}]{Juneau:2005p3499}
Juneau, S., Glazebrook, K., Crampton, D., {et~al.} 2005, The Astrophysical
  Journal, 619, L135

\bibitem[{{Kauffmann}(1995)}]{1995MNRAS.274..153K}
{Kauffmann}, G. 1995, \mnras, 274, 153

\bibitem[{{Kaviraj} {et~al.}(2007){Kaviraj}, {Kirkby}, {Silk}, \&
  {Sarzi}}]{2007MNRAS.382..960K}
{Kaviraj}, S., {Kirkby}, L.~A., {Silk}, J., \& {Sarzi}, M. 2007, \mnras, 382,
  960

\bibitem[{{Khochfar} \& {Silk}(2009)}]{2009MNRAS.397..506K}
{Khochfar}, S. \& {Silk}, J. 2009, \mnras, 397, 506

\bibitem[{Leauthaud {et~al.}(2010)Leauthaud, Finoguenov, Kneib, Taylor, Massey,
  Rhodes, Ilbert, Bundy, Tinker, George, Capak, Koekemoer, Johnston, Zhang,
  Cappelluti, Ellis, Elvis, Giodini, Heymans, F{\`e}vre, Lilly, McCracken,
  Mellier, R{\'e}fr{\'e}gier, Salvato, Scoville, Smoot, Tanaka, Waerbeke, \&
  Wolk}]{Leauthaud:2010p34}
Leauthaud, A., Finoguenov, A., Kneib, J.-P., {et~al.} 2010, The Astrophysical
  Journal, 709, 97

\bibitem[{{Leauthaud} {et~al.}(2011){Leauthaud}, {George}, {Behroozi}, {Bundy},
  {Tinker}, {Wechsler}, {Conroy}, {Finoguenov}, \&
  {Tanaka}}]{2011arXiv1109.0010L}
{Leauthaud}, A., {George}, M.~R., {Behroozi}, P.~S., {et~al.} 2011, ArXiv
  e-prints

\bibitem[{Lilly {et~al.}(2009)Lilly, Brun, Maier, Mainieri, Mignoli, Scodeggio,
  Zamorani, Carollo, Contini, Kneib, F{\`e}vre, Renzini, Bardelli, Bolzonella,
  Bongiorno, Caputi, Coppa, Cucciati, de~la Torre, de~Ravel, Franzetti,
  Garilli, Iovino, Kampczyk, Kovac, Knobel, Lamareille, Borgne, Pello, Peng,
  P{\'e}rez-Montero, Ricciardelli, Silverman, Tanaka, Tasca, Tresse, Vergani,
  Zucca, Ilbert, Salvato, Oesch, Abbas, Bottini, Capak, Cappi, Cassata,
  Cimatti, Elvis, Fumana, Guzzo, Hasinger, Koekemoer, Leauthaud, Maccagni,
  Marinoni, McCracken, Memeo, Meneux, Porciani, Pozzetti, Sanders, Scaramella,
  Scarlata, Scoville, Shopbell, \& Taniguchi}]{Lilly:2009p1622}
Lilly, S.~J., Brun, V.~L., Maier, C., {et~al.} 2009, The Astrophysical Journal
  Supplement, 184, 218

\bibitem[{Longhetti \& Saracco(2009)}]{Longhetti:2009p712}
Longhetti, M. \& Saracco, P. 2009, Monthly Notices of the Royal Astronomical
  Society, 394, 774

\bibitem[{Mandelbaum {et~al.}(2006)Mandelbaum, Seljak, Kauffmann, Hirata, \&
  Brinkmann}]{Mandelbaum:2006p2567}
Mandelbaum, R., Seljak, U., Kauffmann, G., Hirata, C.~M., \& Brinkmann, J.
  2006, Monthly Notices of the Royal Astronomical Society, 368, 715

\bibitem[{McGaugh {et~al.}(2010)McGaugh, Schombert, Blok, \&
  Zagursky}]{Mcgaugh:1p2899}
McGaugh, S., Schombert, J., Blok, W.~D., \& Zagursky, M. 2010, The
  Astrophysical Journal, 708, L14

\bibitem[{McGee {et~al.}(2011)McGee, Balogh, Wilman, Bower, Mulchaey, Parker,
  \& Oemler}]{McGee:2011p2883}
McGee, S.~L., Balogh, M.~L., Wilman, D.~J., {et~al.} 2011, Monthly Notices of
  the Royal Astronomical Society, 413, 996

\bibitem[{Noeske {et~al.}(2007)Noeske, Faber, Weiner, Koo, Primack, Dekel,
  Papovich, Conselice, Floc'h, Rieke, Coil, Lotz, Somerville, \&
  Bundy}]{Noeske:2007p1019}
Noeske, K.~G., Faber, S.~M., Weiner, B.~J., {et~al.} 2007, The Astrophysical
  Journal, 660, L47

\bibitem[{Pannella {et~al.}(2009)Pannella, Gabasch, Goranova, Drory, Hopp,
  Noll, Saglia, Strazzullo, \& Bender}]{Pannella:2009p3946}
Pannella, M., Gabasch, A., Goranova, Y., {et~al.} 2009, The Astrophysical
  Journal, 701, 787

\bibitem[{Patel {et~al.}(2009)Patel, Holden, Kelson, Illingworth, \&
  Franx}]{Patel:2009p1002}
Patel, S.~G., Holden, B.~P., Kelson, D.~D., Illingworth, G.~D., \& Franx, M.
  2009, The Astrophysical Journal Letters, 705, L67

\bibitem[{Peng {et~al.}(2010)Peng, Lilly, Kova{\v c}, Bolzonella, Pozzetti,
  Renzini, Zamorani, Ilbert, Knobel, Iovino, Maier, Cucciati, Tasca, Carollo,
  Silverman, Kampczyk, de~Ravel, Sanders, Scoville, Contini, Mainieri,
  Scodeggio, Kneib, F{\`e}vre, Bardelli, Bongiorno, Caputi, Coppa, de~la Torre,
  Franzetti, Garilli, Lamareille, Borgne, Brun, Mignoli, Montero, Pello,
  Ricciardelli, Tanaka, Tresse, Vergani, Welikala, Zucca, Oesch, Abbas, Barnes,
  Bordoloi, Bottini, Cappi, Cassata, Cimatti, Fumana, Hasinger, Koekemoer,
  Leauthaud, Maccagni, Marinoni, McCracken, Memeo, Meneux, Nair, Porciani,
  Presotto, \& Scaramella}]{Peng:2010p568}
Peng, Y., Lilly, S.~J., Kova{\v c}, K., {et~al.} 2010, The Astrophysical
  Journal, 721, 193

\bibitem[{Polletta {et~al.}(2007)Polletta, Tajer, Maraschi, Trinchieri,
  Lonsdale, Chiappetti, Andreon, Pierre, F{\`e}vre, Zamorani, Maccagni, Garcet,
  Surdej, Franceschini, Alloin, Shupe, Surace, Fang, Rowan-Robinson, Smith, \&
  Tresse}]{Polletta:2007p1833}
Polletta, M., Tajer, M., Maraschi, L., {et~al.} 2007, The Astrophysical
  Journal, 663, 81

\bibitem[{Popesso {et~al.}(2010)Popesso, Biviano, Rodighiero, Baronchelli,
  Salvato, Saintonge, Finoguenov, Magnelli, Gruppioni, Pozzi, Lutz, Elbaz,
  Altieri, Andreani, Aussel, Berta, Capak, Cava, Cimatti, Coia, Daddi,
  Dannerbauer, Dickinson, Dasyra, Fadda, Schreiber, Genzel, Hwang, Kartaltepe,
  Ilbert, Floch, Leiton, Magdis, Nordon, Patel, Poglitsch, Riguccini, Portal,
  Shao, Tacconi, Tomczak, Tran, \& Valtchanov}]{Popesso:10p3436}
Popesso, P., Biviano, A., Rodighiero, G., {et~al.} 2010, eprint arXiv:1110.2946

\bibitem[{Popesso {et~al.}(2005)Popesso, Boehringer, Romaniello, \&
  Voges}]{Popesso:2005p1365}
Popesso, P., Boehringer, H., Romaniello, M., \& Voges, W. 2005, Astronomy and
  Astrophysics, 433, 415

\bibitem[{{Pozzetti} {et~al.}(2010){Pozzetti}, {Bolzonella}, {Zucca},
  {Zamorani}, {Lilly}, {Renzini}, {Moresco}, {Mignoli}, {Cassata}, {Tasca},
  {Lamareille}, {Maier}, {Meneux}, {Halliday}, {Oesch}, {Vergani}, {Caputi},
  {Kova{\v c}}, {Cimatti}, {Cucciati}, {Iovino}, {Peng}, {Carollo}, {Contini},
  {Kneib}, {Le F{\'e}vre}, {Mainieri}, {Scodeggio}, {Bardelli}, {Bongiorno},
  {Coppa}, {de la Torre}, {de Ravel}, {Franzetti}, {Garilli}, {Kampczyk},
  {Knobel}, {Le Borgne}, {Le Brun}, {Pell{\`o}}, {Perez Montero},
  {Ricciardelli}, {Silverman}, {Tanaka}, {Tresse}, {Abbas}, {Bottini}, {Cappi},
  {Guzzo}, {Koekemoer}, {Leauthaud}, {Maccagni}, {Marinoni}, {McCracken},
  {Memeo}, {Porciani}, {Scaramella}, {Scarlata}, \&
  {Scoville}}]{2010A&A...523A..13P}
{Pozzetti}, L., {Bolzonella}, M., {Zucca}, E., {et~al.} 2010, \aap, 523, A13+

\bibitem[{Rettura {et~al.}(2011)Rettura, Mei, Stanford, Raichoor, Moran,
  Holden, Rosati, Ellis, Nakata, Nonino, Treu, Blakeslee, Demarco, Eisenhardt,
  Ford, Fosbury, Illingworth, Huertas-Company, Jee, Kodama, Postman, Tanaka, \&
  White}]{Rettura:2011p3936}
Rettura, A., Mei, S., Stanford, S.~A., {et~al.} 2011, The Astrophysical
  Journal, 732, 94

\bibitem[{Salpeter(1955)}]{Salpeter:1955p1869}
Salpeter, E.~E. 1955, Astrophysical Journal, 121, 161

\bibitem[{Schechter(1976)}]{Schechter:1976p1608}
Schechter, P. 1976, Astrophys. J., 203, 297

\bibitem[{Scoville {et~al.}(2007)Scoville, Aussel, Brusa, Capak, Carollo,
  Elvis, Giavalisco, Guzzo, Hasinger, Impey, Kneib, LeFevre, Lilly, Mobasher,
  Renzini, Rich, Sanders, Schinnerer, Schminovich, Shopbell, Taniguchi, \&
  Tyson}]{Scoville:2007p1888}
Scoville, N., Aussel, H., Brusa, M., {et~al.} 2007, The Astrophysical Journal
  Supplement Series, 172, 1

\bibitem[{{Strateva} {et~al.}(2001){Strateva}, {Ivezi{\'c}}, {Knapp},
  {Narayanan}, {Strauss}, {Gunn}, {Lupton}, {Schlegel}, {Bahcall}, {Brinkmann},
  {Brunner}, {Budav{\'a}ri}, {Csabai}, {Castander}, {Doi}, {Fukugita}, {Gy{\H
  o}ry}, {Hamabe}, {Hennessy}, {Ichikawa}, {Kunszt}, {Lamb}, {McKay},
  {Okamura}, {Racusin}, {Sekiguchi}, {Schneider}, {Shimasaku}, \&
  {York}}]{2001AJ....122.1861S}
{Strateva}, I., {Ivezi{\'c}}, {\v Z}., {Knapp}, G.~R., {et~al.} 2001, \aj, 122,
  1861

\bibitem[{Trentham \& Tully(2002)}]{Trentham:2002p4697}
Trentham, N. \& Tully, R.~B. 2002, Monthly Notice of the Royal Astronomical
  Society, 335, 712

\bibitem[{van~den Bosch {et~al.}(2008)van~den Bosch, Aquino, Yang, Mo,
  Pasquali, McIntosh, Weinmann, \& Kang}]{vandenBosch:2008p3931}
van~den Bosch, F.~C., Aquino, D., Yang, X., {et~al.} 2008, Monthly Notices of
  the Royal Astronomical Society, 387, 79

\bibitem[{van Dokkum(2005)}]{vanDokkum:2005p4410}
van Dokkum, P.~G. 2005, The Astronomical Journal, 130, 2647

\bibitem[{van Uitert {et~al.}(2011)van Uitert, Hoekstra, Velander, Gilbank,
  Gladders, \& Yee}]{vanUitert:2011p2963}
van Uitert, E., Hoekstra, H., Velander, M., {et~al.} 2011, eprint arXiv, 1107,
  4093

\bibitem[{Vulcani {et~al.}(2010{\natexlab{a}})Vulcani, Poggianti,
  Arag{\'o}n-Salamanca, Fasano, Rudnick, Valentinuzzi, Dressler, Bettoni, Cava,
  D'Onofrio, Fritz, Moretti, Omizzolo, \& Varela}]{Vulcani:2010p919}
Vulcani, B., Poggianti, B.~M., Arag{\'o}n-Salamanca, A., {et~al.}
  2010{\natexlab{a}}, eprint arXiv, 1010, 4442

\bibitem[{Vulcani {et~al.}(2010{\natexlab{b}})Vulcani, Poggianti, Finn,
  Rudnick, Desai, \& Bamford}]{Vulcani:2010p1322}
Vulcani, B., Poggianti, B.~M., Finn, R.~A., {et~al.} 2010{\natexlab{b}}, The
  Astrophysical Journal Letters, 710, L1

\bibitem[{{Weinmann} {et~al.}(2011){Weinmann}, {Lisker}, {Guo}, {Meyer}, \&
  {Janz}}]{2011arXiv1105.0674W}
{Weinmann}, S.~M., {Lisker}, T., {Guo}, Q., {Meyer}, H.~T., \& {Janz}, J. 2011,
  ArXiv e-prints

\bibitem[{Wetzel {et~al.}(2011)Wetzel, Tinker, \& Conroy}]{Wetzel:2011p3934}
Wetzel, A.~R., Tinker, J.~L., \& Conroy, C. 2011, eprint arXiv, 1107, 5311

\bibitem[{White \& Rees(1978)}]{White:1978p1621}
White, S. D.~M. \& Rees, M.~J. 1978, Royal Astronomical Society, 183, 341

\bibitem[{{Wilman} {et~al.}(2005){Wilman}, {Balogh}, {Bower}, {Mulchaey},
  {Oemler}, {Carlberg}, {Morris}, \& {Whitaker}}]{2005MNRAS.358...71W}
{Wilman}, D.~J., {Balogh}, M.~L., {Bower}, R.~G., {et~al.} 2005, \mnras, 358,
  71

\bibitem[{Wilson {et~al.}(1997)Wilson, Smail, Ellis, \&
  Couch}]{Wilson:1997p4567}
Wilson, G., Smail, I., Ellis, R.~S., \& Couch, W.~J. 1997, Monthly Notices of
  the Royal Astronomical Society, 284, 915

\bibitem[{Yang {et~al.}(2009)Yang, Mo, \& van~den Bosch}]{Yang:2009p343}
Yang, X., Mo, H.~J., \& van~den Bosch, F.~C. 2009, The Astrophysical Journal,
  695, 900

\bibitem[{{Zhang} {et~al.}(2011){Zhang}, {Lagan{\'a}}, {Pierini}, {Puchwein},
  {Schneider}, \& {Reiprich}}]{2011arXiv1109.0390Z}
{Zhang}, Y.-Y., {Lagan{\'a}}, T.~F., {Pierini}, D., {et~al.} 2011, ArXiv
  e-prints

\end{thebibliography}
\appendix
\onecolumn
\renewcommand{\arraystretch}{2}
\section{Galaxy Stellar Mass Distributions}\label{table_gsmf}
\begin{center}
\longtab{1}{
\begin{longtable}{l||lll|lll|lll}
\caption[Galaxy  Stellar Mass Distribution of the COSMOS Groups and Field.]{Galaxy  Stellar Mass Distribution of the COSMOS Groups and Field. We list the GSMF for all the galaxies  (N$_\mathrm{SFG}$) and separately for passive (N$_\mathrm{passive}$) and star forming (N$_\mathrm{SFG}$) galaxies in the different environments. The groups GSMF is background subtracted.}\\
\hline
\hline
Log(M)& N$_\mathrm{SFG}$ & N$_\mathrm{passive}$& N$_\mathrm{all}$& N$_\mathrm{SFG}$ & N$_\mathrm{passive}$& N$_\mathrm{all}$& N$_\mathrm{SFG}$ & N$_\mathrm{passive}$& N$_\mathrm{all}$\\
\hline 
\endfirsthead
\setlength{\tabcolsep}{2pt}\\
\hline
\hline
Log(M)& N$_\mathrm{SFG}$ & N$_\mathrm{passive}$& N$_\mathrm{all}$& N$_\mathrm{SFG}$ & N$_\mathrm{passive}$& N$_\mathrm{all}$& N$_\mathrm{SFG}$ & N$_\mathrm{passive}$& N$_\mathrm{all}$\\
\hline 
\multicolumn{1}{c}{} &\multicolumn{3}{c}{LOW MASS GROUPS}& \multicolumn{3}{c}{HIGH MASS GROUPS} & \multicolumn{3}{c}{FIELD} \\
\hline 
\endhead
\hline
\endfoot 
\hline
\hline 
\multicolumn{10}{c}{z=0.2--0.4}\\
\hline 
 8.125&161.92$^{70.46}_{178.23}$&13.07$^{6.16}_{21.73}$&174.99$^{84.09}_{191.78}$&41.30$^{10.57}_{51.25}$&5.93$^{0.67}_{13.89}$&47.24$^{11.08}_{58.27}$&3411$^{2319.44}_{3469.40}$&99$^{80.16}_{108.95}$&3510$^{2482.29}_{3569.25}$\\
 8.375&186.95$^{135.96}_{203.68}$&37.75$^{20.14}_{46.94}$&224.70$^{176.81}_{246.31}$&72.67$^{44.62}_{84.56}$&17.73$^{7.34}_{24.58}$&90.40$^{58.34}_{102.29}$&3005$^{2852.79}_{3059.82}$&211$^{194.42}_{232.61}$&3216$^{3008.12}_{3272.71}$\\
 8.625&155.90$^{136.03}_{195.93}$&55.62$^{36.26}_{69.99}$&211.52$^{182.84}_{257.84}$&61.28$^{47.84}_{77.93}$&23.87$^{11.30}_{36.58}$&85.15$^{65.90}_{106.09}$&2298$^{2249.90}_{2399.97}$&384$^{294.82}_{403.60}$&2682$^{2628.08}_{2756.10}$\\
 8.875&149.53$^{121.63}_{179.50}$&44.41$^{31.59}_{60.71}$&193.93$^{166.74}_{221.38}$&53.31$^{39.73}_{69.50}$&34.88$^{22.75}_{47.22}$&88.19$^{73.69}_{109.94}$&1737$^{1695.32}_{1893.65}$&290$^{262.96}_{312.05}$&2027$^{1981.98}_{2157.98}$\\
 9.125& 90.28$^{75.63}_{122.42}$&42.25$^{30.45}_{62.88}$&132.54$^{118.19}_{163.63}$&41.22$^{28.29}_{55.95}$&33.91$^{23.47}_{45.36}$&75.13$^{60.07}_{99.58}$&1374$^{1319.47}_{1415.21}$&194$^{180.07}_{226.17}$&1568$^{1524.50}_{1632.57}$\\
 9.375& 82.75$^{65.96}_{101.26}$&49.18$^{34.76}_{62.21}$&131.93$^{103.81}_{156.72}$&37.38$^{22.31}_{50.21}$&28.61$^{19.99}_{38.95}$&65.99$^{52.18}_{81.57}$&  987$^{955.58}_{1066.47}$&129$^{117.64}_{153.76}$&1116$^{1082.59}_{1215.76}$\\
 9.625&  53.91$^{43.12}_{76.25}$&38.80$^{26.48}_{51.33}$& 92.71$^{76.43}_{118.21}$&27.25$^{18.25}_{39.50}$&29.84$^{19.67}_{43.55}$&57.09$^{43.71}_{69.32}$&   813$^{784.47}_{854.39}$&108$^{97.18}_{123.87}$&   921$^{887.98}_{962.29}$\\
 9.875&  46.27$^{27.11}_{58.42}$&32.53$^{22.83}_{45.62}$&  78.81$^{60.73}_{93.74}$&21.20$^{14.48}_{32.23}$&29.74$^{18.77}_{41.74}$&50.94$^{41.65}_{68.30}$&   632$^{591.31}_{658.38}$&117$^{105.47}_{137.15}$&   749$^{718.70}_{782.32}$\\
 10.125&  49.89$^{34.24}_{68.34}$&23.76$^{16.54}_{38.24}$&  73.66$^{57.07}_{91.28}$& 16.88$^{6.38}_{23.01}$&31.46$^{17.84}_{39.91}$&48.34$^{32.16}_{58.74}$&   476$^{454.18}_{526.91}$&143$^{128.03}_{158.59}$&   619$^{593.41}_{659.53}$\\
 10.375&  29.90$^{21.98}_{42.49}$&32.70$^{22.14}_{42.77}$&  62.60$^{51.14}_{82.36}$& 11.24$^{5.04}_{20.73}$&29.07$^{18.10}_{36.36}$&40.32$^{28.56}_{52.30}$&   442$^{407.78}_{464.49}$&179$^{159.64}_{198.36}$&   621$^{581.23}_{646.63}$\\
 10.625&  30.26$^{21.67}_{40.40}$&31.78$^{19.63}_{41.18}$&  62.04$^{41.68}_{72.34}$& 16.83$^{8.11}_{23.14}$&11.74$^{7.26}_{24.68}$&28.57$^{20.33}_{41.55}$&   295$^{273.46}_{320.61}$&210$^{188.41}_{224.63}$&   505$^{475.57}_{533.18}$\\
 10.875&  20.87$^{12.51}_{28.24}$&25.18$^{17.87}_{37.10}$&  46.05$^{35.95}_{57.73}$&   3.14$^{0.01}_{8.39}$&16.61$^{8.03}_{22.31}$&19.75$^{10.84}_{26.35}$&   173$^{158.54}_{189.52}$&129$^{114.51}_{150.28}$&   302$^{283.27}_{330.04}$\\
 11.125&    4.16$^{0.78}_{10.33}$&24.84$^{15.67}_{33.52}$&  29.00$^{20.18}_{38.78}$&   1.33$^{0.01}_{4.18}$&7.21$^{2.54}_{13.59}$&  8.55$^{4.15}_{14.41}$&      62$^{54.13}_{73.96}$& 73$^{63.57}_{86.15}$&   135$^{123.34}_{151.00}$\\
 11.375&     2.85$^{0.01}_{6.41}$&12.29$^{6.38}_{18.47}$&   15.14$^{8.25}_{21.62}$&   0.95$^{0.01}_{3.27}$& 2.74$^{0.01}_{6.94}$&   3.69$^{0.24}_{8.06}$&        5$^{2.00}_{10.48}$& 24$^{17.00}_{30.32}$&      29$^{20.94}_{35.16}$\\
 11.625&     1.00$^{0.01}_{3.32}$& -0.06$^{0.01}_{2.68}$&     0.94$^{0.01}_{4.01}$&   0.00$^{0.01}_{2.12}$& 1.98$^{0.01}_{5.31}$&   1.98$^{0.01}_{5.31}$&         0$^{0.01}_{0.00}$&    2$^{0.01}_{3.41}$&         2$^{0.01}_{4.45}$\\

\hline
\hline
\multicolumn{10}{c}{z=0.4--0.6}\\
\hline
 8.125&84.52$^{55.37}_{96.22}$&-0.03$^{0.01}_{4.39}$&84.50$^{50.66}_{97.02}$&       - &       - &       - &2632$^{2115.45}_{2683.30}$&5$^{2.55}_{20.17}$&2637$^{2141.33}_{2688.35}$\\
 8.375&142.12$^{103.79}_{154.97}$& 1.80$^{0.01}_{8.37}$&143.92$^{108.33}_{161.53}$&       - &       - &       - &3979$^{3437.31}_{4042.08}$&36$^{30.00}_{63.66}$&4015$^{3407.69}_{4078.36}$\\
 8.625&161.40$^{116.28}_{174.93}$&17.26$^{5.19}_{23.38}$&178.66$^{133.35}_{193.15}$&       - &       - &       - &3928$^{3611.73}_{3990.67}$&134$^{96.18}_{145.58}$&4062$^{3759.22}_{4125.73}$\\
 8.875&124.73$^{99.72}_{141.64}$&20.88$^{11.22}_{30.74}$&145.60$^{108.35}_{166.09}$&       - &       - &       - &2959$^{2904.60}_{3152.79}$&204$^{143.30}_{218.28}$&3163$^{3103.95}_{3305.56}$\\
 9.125&87.44$^{76.76}_{116.61}$&22.15$^{12.43}_{30.54}$&109.59$^{96.51}_{140.64}$&       - &       - &       - &2101$^{2055.16}_{2264.55}$&155$^{137.71}_{173.73}$&2256$^{2208.50}_{2442.16}$\\
 9.375&69.84$^{58.58}_{86.98}$&16.45$^{9.48}_{24.40}$&86.29$^{73.14}_{108.14}$&       - &       - &       - &1484$^{1445.48}_{1597.72}$&100$^{89.23}_{131.62}$&1584$^{1544.20}_{1699.10}$\\
 9.625&53.10$^{39.28}_{72.22}$&22.53$^{12.82}_{31.67}$&75.64$^{61.67}_{92.10}$&       - &       - &       - &1072$^{1039.26}_{1153.84}$&85$^{75.78}_{107.93}$&1157$^{1122.99}_{1242.09}$\\
 9.875&27.66$^{19.56}_{43.19}$&16.43$^{10.72}_{27.71}$&44.08$^{32.98}_{61.11}$&       - &       - &       - &   790$^{760.78}_{829.67}$&104$^{91.04}_{115.83}$&   894$^{864.10}_{953.98}$\\
 10.125&41.12$^{28.67}_{51.57}$&18.21$^{11.08}_{26.32}$&59.33$^{45.31}_{72.09}$&       - &       - &       - &   706$^{665.93}_{735.15}$&143$^{130.04}_{162.97}$&   849$^{806.46}_{895.31}$\\
 10.375&26.84$^{19.47}_{40.51}$&23.84$^{15.55}_{32.38}$&50.68$^{39.28}_{69.67}$&       - &       - &       - &   575$^{547.23}_{614.19}$&211$^{187.08}_{227.58}$&   786$^{750.97}_{824.92}$\\
 10.625&34.43$^{21.50}_{41.85}$&24.73$^{16.38}_{33.34}$&59.17$^{40.99}_{69.97}$&       - &       - &       - &   467$^{426.71}_{491.25}$&230$^{210.03}_{248.73}$&   697$^{663.27}_{728.95}$\\
 10.875&20.32$^{14.91}_{32.74}$&19.89$^{13.25}_{28.90}$&40.21$^{32.68}_{53.59}$&       - &       - &       - &   306$^{288.51}_{334.11}$&201$^{180.36}_{216.39}$&   507$^{480.49}_{534.62}$\\
 11.125&  5.38$^{0.01}_{10.06}$&16.30$^{8.59}_{22.32}$&21.68$^{10.26}_{28.28}$&       - &       - &       - &   113$^{102.18}_{135.65}$&127$^{115.55}_{144.20}$&   240$^{224.00}_{272.00}$\\
 11.375&  -0.08$^{0.01}_{2.04}$&8.79$^{4.53}_{15.27}$&  8.71$^{4.82}_{16.83}$&       - &       - &       - &      14$^{10.26}_{23.75}$&39$^{30.34}_{49.15}$&      53$^{45.45}_{67.90}$\\
 11.625&   0.00$^{0.01}_{1.87}$& 2.97$^{0.01}_{5.90}$&   2.97$^{0.01}_{5.90}$&       - &       - &       - &         0$^{0.01}_{2.00}$& 6$^{3.35}_{9.16}$&         6$^{3.35}_{9.87}$\\

\hline
\multicolumn{10}{c}{z=0.6--0.8}\\
\hline
 8.125&54.66$^{32.90}_{63.59}$&-0.04$^{0.01}_{3.50}$&54.62$^{28.19}_{63.77}$&       - &       - &       - &1467$^{1416.44}_{1540.73}$&7$^{0.44}_{9.83}$&1474$^{1418.56}_{1524.63}$\\
 8.375&87.06$^{66.31}_{107.75}$& 0.98$^{0.01}_{5.61}$&88.04$^{73.64}_{114.13}$&       - &       - &       - &3385$^{3090.20}_{3443.18}$&4$^{1.76}_{11.28}$&3389$^{3087.33}_{3447.22}$\\
 8.625&152.23$^{115.64}_{166.83}$& 3.95$^{0.01}_{7.28}$&156.18$^{112.96}_{169.96}$&       - &       - &       - &4949$^{4469.81}_{5019.35}$&10$^{6.84}_{20.49}$&4959$^{4470.89}_{5029.42}$\\
 8.875&159.03$^{130.53}_{183.22}$&2.79$^{0.01}_{10.39}$&161.82$^{132.39}_{183.67}$&       - &       - &       - &5188$^{4880.45}_{5260.03}$&41$^{27.40}_{53.73}$&5229$^{4910.68}_{5301.31}$\\
 9.125&149.41$^{125.51}_{168.58}$&11.46$^{1.52}_{16.45}$&160.87$^{135.89}_{185.86}$&       - &       - &       - &4313$^{4242.37}_{4472.18}$&108$^{74.35}_{118.39}$&4421$^{4329.40}_{4552.97}$\\
 9.375&125.47$^{105.55}_{146.24}$&15.43$^{5.27}_{23.30}$&140.91$^{121.36}_{166.25}$&       - &       - &       - &3102$^{3046.30}_{3338.65}$&113$^{83.05}_{123.63}$&3215$^{3158.30}_{3433.49}$\\
 9.625&107.04$^{86.87}_{126.39}$&14.51$^{6.95}_{23.09}$&121.55$^{101.83}_{147.29}$&       - &       - &       - &2189$^{2142.21}_{2330.93}$&98$^{88.10}_{120.32}$&2287$^{2239.18}_{2415.25}$\\
 9.875&69.22$^{57.45}_{89.73}$&18.23$^{9.58}_{28.75}$&87.45$^{72.96}_{105.96}$&       - &       - &       - &1554$^{1514.58}_{1655.01}$&154$^{126.98}_{170.58}$&1708$^{1666.67}_{1810.68}$\\
 10.125&57.90$^{44.65}_{75.42}$&22.97$^{13.20}_{34.62}$&80.88$^{66.68}_{99.14}$&       - &       - &       - &1218$^{1181.96}_{1266.72}$&205$^{186.94}_{244.67}$&1423$^{1385.28}_{1488.05}$\\
 10.375&52.04$^{41.22}_{65.23}$&31.53$^{21.25}_{42.03}$&83.57$^{68.58}_{98.79}$&       - &       - &       - &  991$^{955.69}_{1034.49}$&294$^{266.11}_{316.14}$&1285$^{1240.12}_{1330.49}$\\
 10.625&36.36$^{24.94}_{50.45}$&33.39$^{21.39}_{45.58}$&69.75$^{56.06}_{90.20}$&       - &       - &       - &   727$^{699.14}_{776.07}$&322$^{300.95}_{346.04}$&1049$^{1016.61}_{1124.32}$\\
 10.875&26.28$^{17.73}_{35.10}$&23.46$^{15.86}_{36.02}$&49.74$^{38.39}_{64.39}$&       - &       - &       - &   543$^{502.60}_{568.36}$&308$^{280.63}_{327.72}$&   851$^{785.18}_{880.17}$\\
 11.125&27.88$^{18.60}_{36.70}$&33.96$^{21.97}_{41.01}$&61.85$^{46.21}_{71.12}$&       - &       - &       - &   223$^{208.07}_{255.62}$&207$^{186.93}_{222.59}$&   430$^{409.24}_{485.98}$\\
 11.375&   2.78$^{0.01}_{6.98}$&15.49$^{8.61}_{23.36}$&18.27$^{10.41}_{27.95}$&       - &       - &       - &      44$^{37.07}_{55.18}$&101$^{90.90}_{119.89}$&   145$^{132.79}_{168.35}$\\
 11.625&  -0.03$^{0.01}_{2.09}$& 2.95$^{0.38}_{7.91}$&  2.92$^{0.36}_{10.52}$&       - &       - &       - &         5$^{2.00}_{7.45}$&10$^{6.84}_{20.49}$&      15$^{11.13}_{28.56}$\\

\hline
\multicolumn{10}{c}{z=0.8--1.0}\\
\hline
 8.125&7.25$^{0.26}_{14.45}$&-0.01$^{0.01}_{2.72}$&7.24$^{1.09}_{15.29}$&10.98$^{0.14}_{15.99}$&0.99$^{0.01}_{4.05}$&11.97$^{2.87}_{17.10}$&353$^{334.21}_{510.13}$&5$^{2.00}_{12.35}$&358$^{339.08}_{513.16}$\\
 8.375&21.34$^{11.57}_{30.55}$& 0.96$^{0.01}_{3.49}$&22.30$^{9.91}_{34.87}$&12.39$^{7.25}_{31.10}$&-0.05$^{0.01}_{3.48}$&12.33$^{7.20}_{34.92}$&1256$^{1220.56}_{1441.42}$&18$^{4.33}_{22.24}$&1274$^{1238.31}_{1434.03}$\\
 8.625&46.35$^{31.08}_{62.64}$& 1.95$^{0.01}_{4.61}$&48.30$^{33.80}_{64.66}$&59.95$^{37.09}_{69.52}$&1.93$^{0.01}_{5.94}$&61.88$^{38.98}_{71.36}$&3144$^{2877.04}_{3200.07}$&24$^{9.17}_{28.90}$&3168$^{2926.35}_{3224.28}$\\
 8.875&91.10$^{62.24}_{103.39}$& 0.98$^{0.01}_{3.51}$&92.08$^{65.99}_{103.40}$&75.17$^{58.55}_{98.67}$&1.98$^{0.01}_{5.30}$&77.15$^{56.34}_{94.73}$&5152$^{4507.99}_{5223.78}$&8$^{5.17}_{21.30}$&5160$^{4519.96}_{5231.83}$\\
 9.125&94.33$^{65.44}_{114.18}$&-0.06$^{0.01}_{5.28}$&94.27$^{67.25}_{117.64}$&94.49$^{67.39}_{113.59}$&-0.07$^{0.01}_{5.26}$&94.41$^{72.72}_{111.09}$&5044$^{4879.84}_{5121.49}$&26$^{15.66}_{35.49}$&5070$^{4922.65}_{5141.20}$\\
 9.375&96.04$^{69.95}_{111.08}$& 3.90$^{0.01}_{7.66}$&99.94$^{77.38}_{113.73}$&93.82$^{73.08}_{110.42}$&2.86$^{0.01}_{6.42}$&96.69$^{69.59}_{109.26}$&4232$^{4166.75}_{4409.36}$&47$^{31.41}_{53.86}$&4279$^{4213.59}_{4446.32}$\\
 9.625&68.00$^{55.02}_{85.87}$& 0.84$^{0.01}_{5.47}$&68.85$^{53.60}_{89.32}$&80.48$^{68.13}_{101.34}$&5.79$^{1.38}_{11.17}$&86.27$^{71.07}_{105.60}$&3307$^{3249.49}_{3489.31}$&74$^{61.55}_{90.43}$&3381$^{3322.85}_{3521.59}$\\
 9.875&65.64$^{52.83}_{81.73}$&6.68$^{1.43}_{12.19}$&72.32$^{53.70}_{94.63}$&50.71$^{41.01}_{68.91}$&17.57$^{7.29}_{23.26}$&68.28$^{55.93}_{89.64}$&2534$^{2483.66}_{2695.07}$&149$^{130.43}_{172.43}$&2683$^{2631.20}_{2808.22}$\\
 10.125&55.63$^{47.13}_{71.33}$&9.41$^{3.82}_{16.78}$&65.05$^{52.96}_{93.64}$&52.06$^{37.47}_{65.30}$&21.20$^{11.54}_{30.27}$&73.26$^{57.87}_{87.40}$&2063$^{2010.66}_{2129.81}$&278$^{251.19}_{299.14}$&2341$^{2285.10}_{2400.72}$\\
 10.375&53.47$^{37.22}_{64.56}$&21.13$^{10.63}_{26.90}$&74.60$^{48.76}_{85.30}$&48.20$^{35.40}_{60.46}$&30.81$^{23.30}_{42.86}$&79.01$^{62.67}_{96.37}$&1669$^{1625.13}_{1748.33}$&412$^{386.15}_{437.24}$&2081$^{2032.32}_{2173.96}$\\
 10.625&34.25$^{26.37}_{49.08}$&21.97$^{13.82}_{30.37}$&56.22$^{45.25}_{74.48}$&40.26$^{27.86}_{49.43}$&34.60$^{25.60}_{46.84}$&74.86$^{56.18}_{85.62}$&1300$^{1243.11}_{1336.06}$&487$^{463.53}_{511.66}$&1787$^{1708.62}_{1832.56}$\\
 10.875&29.29$^{18.32}_{37.04}$&22.01$^{13.86}_{31.14}$&51.29$^{38.46}_{61.07}$&26.67$^{18.79}_{39.43}$&22.65$^{15.13}_{31.85}$&49.32$^{39.90}_{63.93}$&809$^{780.56}_{888.28}$&470$^{432.99}_{496.36}$&1279$^{1241.87}_{1338.86}$\\
 11.125&13.21$^{7.21}_{20.17}$&11.44$^{6.78}_{18.98}$&24.65$^{18.50}_{33.99}$&10.93$^{6.73}_{19.29}$&15.23$^{6.79}_{21.15}$&26.17$^{18.35}_{37.19}$&371$^{351.71}_{403.36}$&266$^{249.21}_{288.85}$&637$^{611.68}_{703.94}$\\
 11.375&5.82$^{2.02}_{10.51}$&11.81$^{6.00}_{16.49}$&17.63$^{8.21}_{23.32}$&  4.76$^{1.13}_{8.70}$&9.74$^{4.14}_{14.96}$&14.50$^{9.00}_{20.88}$&  84$^{74.83}_{109.69}$&91$^{78.55}_{114.98}$&175$^{161.77}_{212.42}$\\
 11.625&-0.01$^{0.01}_{2.11}$& 0.98$^{0.01}_{4.04}$& 0.97$^{0.01}_{4.03}$&  1.99$^{0.01}_{4.83}$&2.97$^{0.01}_{7.17}$&  4.96$^{1.33}_{9.49}$&     4$^{2.00}_{13.22}$&11$^{7.13}_{19.66}$&   15$^{11.13}_{26.66}$\\

\hline
\hline

\end{longtable}%
}
\end{center}
\end{document}